\documentclass[aps,pra,twocolumn,showpacs, superscriptaddress]{revtex4-1}

\usepackage{graphicx}
\usepackage{color}
\usepackage[usenames,dvipsnames]{xcolor}
\usepackage{amsfonts}
\usepackage{bm}
\usepackage{amsmath}

\newcommand{\nc}{\newcommand}
\nc{\eq}{\begin{equation}}
\nc{\eeq}{\end{equation}}
\nc{\eqa}{\begin{eqnarray}}
\nc{\eeqa}{\end{eqnarray}}
\nc{\ar}{\begin{array}}
\nc{\ear}{\end{array}}
\nc{\bfig}{\begin{figure}}
\nc{\efig}{\end{figure}}
\nc{\dg}{\dagger}
\nc{\eps}{\frac{\epsilon}{2}}
\nc{\juuri}{\sqrt{\Omega^2+(\eps)^2}}
\nc{\sx}{\sigma_x}
\nc{\sy}{\sigma_y}
\nc{\sz}{\sigma_z}
\nc{\spl}{\sigma_+}
\nc{\sm}{\sigma_-}
\nc{\Sx}{\bar{\sigma}_x}
\nc{\Sy}{\bar{\sigma}_y}
\nc{\Sz}{\bar{\sigma}_z}
\nc{\Spl}{\bar{\sigma}_+}
\nc{\Sm}{\bar{\sigma}_-}
\nc{\nn}{\nonumber}
\nc{\noi}{\noindent}
\nc{\omt}{\tilde{\omega}}
\nc{\Somt}{S(\omt)}
\nc{\Somtd}{S^{\dg}(\omt)}
\nc{\got}{\gamma_{\omega}(t)}
\nc{\gmot}{\gamma_{-\omega}(t)}
\nc{\po}{\mathcal{P}}
\nc{\qo}{\mathcal{Q}}
\nc{\adg}{a^{\dg}}
\nc{\gammat}{\tilde{\gamma}}
\nc{\Q}{$\mathcal{Q }$}
\nc{\C}{$\mathcal{C }$}
\nc{\N}{\mathcal{N}}
\nc{\kvec}{\mathbf{k}}

\def\bra#1{\mathinner{\langle{#1}|}}
\def\ket#1{\mathinner{|{#1}\rangle}}

\def\oper{{\mathchoice{\rm 1\mskip-4mu l}{\rm 1\mskip-4mu l}{\rm 1\mskip-4.5mu l}{\rm 1\mskip-5mu l}}}

\begin{document}

%Title of paper
\title{Comparative study of non-Markovianity measures in exactly solvable one and two qubit models}

\author{Carole Addis}
%\email{ca99@hw.ac.uk}
\thanks{These two authors contributed equally to this work.}
\affiliation{SUPA, EPS/Physics, Heriot-Watt University, Edinburgh, EH14 4AS, UK}

\author{Bogna Bylicka}
\thanks{These two authors contributed equally to this work.}
\affiliation{Institute of Physics, Faculty of Physics, Astronomy and Informatics, Nicolaus Copernicus University, Grudziadzka 5, 87-100 Toru\'{n}, Poland}

\author{ Dariusz Chru\'sci\'{n}ski}
\affiliation{Institute of Physics, Faculty of Physics, Astronomy and Informatics, Nicolaus Copernicus University, Grudziadzka 5, 87-100 Toru\'{n}, Poland}

\author{Sabrina Maniscalco} 
%\email{s.maniscalco@hw.ac.uk} 
%\affiliation{SUPA, EPS/Physics, Heriot-Watt University, Edinburgh, EH14 4AS, UK}
\affiliation{Turku Center for Quantum Physics, Department of Physics and Astronomy, University of Turku, FIN-20014 Turku, Finland}

\date{\today}

\begin{abstract}
In this paper we present a detailed critical study of several recently proposed non-Markovianity measures. We analyse their properties for single qubit and two-qubit systems in both pure-dephasing and dissipative scenarios. More specifically we investigate and compare their computability, their physical meaning, their Markovian to non-Markovian crossover, and their additivity properties with respect to the number of qubits. The bottom-up approach that we pursue is aimed at identifying  similarities and differences in the behavior of non-Markovianity indicators in several paradigmatic open system models. This in turn allows us to infer the leading traits of the variegated phenomenon known as non-Markovian dynamics.

\end{abstract}


\maketitle

\section{Introduction}
The last decade has seen a renewed interest in theoretical and experimental investigations on fundamental studies of open quantum systems.
The reasons for such interest are manifold. On the one hand,  we have witnessed a tremendous advance in the development of quantum technologies, stemming from the ability to coherently control in a robust and efficient manner the dynamics of an ever increasing number of particles. Quantum technologies need to be scalable to reach the market, which entails understanding and minimizing environment-induced decoherence effects in order to achieve the required thresholds for error correction. On the other hand the environment itself has been proven to be experimentally controllable and modifiable. Nowadays reservoir engineering techniques are used for both minimizing the effects of environmental noise and as testbeds of theoretical models. 

All recent investigations on open quantum systems dynamics highlight the existence of two different classes of dynamical behavior  known as Markovian and non-Markovian regimes. Historically, in the quantum domain Markovian dynamics has been associated to the semigroup property of the dynamical map describing the system evolution. If one thinks in terms of a microscopic model of system, environment and interaction, a Markovian description of the open system requires a number of assumptions, such as system-reservoir weak coupling, and leads to a master equation in the so-called Lindblad form \cite{Lindblad, GKS}. In certain scenarios, however, such approximations are not justified and one needs to go beyond perturbation theory. It is clear that, due to the general complexity of the problem to be studied, exact solutions exist only for simple open quantum systems models such as the well-known Jaynes-Cummings model \cite{Barry}, the quantum Brownian motion model \cite{QBM}, and certain pure dephasing models \cite{puremodel1}-\cite{puremodel3}. Despite the fact that these models are often idealized versions of what can be implemented in current experiments, it is undoubtedly very important to fully understand and study these systems and compare the theoretical predictions with experimental implementations. 

One of the first features that emerges from the analysis of exact models is that memory effects, usually associated to re-coherence and information backflow, are not connected to the semigroup property of the dynamical map, but are rather associated to the more general property known as divisibility, as discussed for example in Ref. \cite{quantumjumps}-\cite{RHP}. In this spirit, a non-Markovianity measure quantifying the deviation from divisibility has been proposed in Ref. \cite{RHP}. Other measures (and corresponding definitions) are based on the behavior of quantities such as distinguishability between quantum states, as measured, e.g., by trace distance \cite{BLP, NMPure1i} or fidelity \cite{fid}, quantum mutual information between initial and final state \cite{LFS}, channel capacities \cite{Bogna}, Fisher information \cite{fish}, and the volume of accessible physical states of a system \cite{plastina}.

What one defines as Markovian or non-Markovian dynamics is, in a sense, a question of semantics. It is tautologic to say that, in general, different definitions and corresponding measures of non-Markovianity do not coincide. We prefer to follow a more pragmatic approach. We will not insist on the concept of \lq \lq the best\rq \rq definition of non-Markovianity  but we rather look at different measures as descriptions of different properties of the open quantum systems.

Our study will encompass the Rivas, Huelga, Pleanio (RHP) divisibility measure of  \cite{RHP}, the Breuer, Laine, Piilo (BLP) distinguishability measure  \cite{BLP}, the Luo, Fu, Song (LFS) coherent information measure \cite{LFS}, and the Bylicka, Chru\'sci\'{n}ski, Maniscalco (BCM)  channel capacity measures \cite{Bogna}. We will first consider the case of one and two qubits immersed in independent and common purely dephasing environments. We will then extend our analysis to the study of dissipative environments, again for single qubit and two qubits. 

The paper is structured as follows. We begin by reviewing both the definitions and properties of the non-Markovianity measures used in the paper (Sec. II) and their properties and physical meaning (Sec. III). In Secs. IV and V we present the results of the dynamics of single qubit and two qubits interacting with pure dephasing and amplitude damping environment, respectively. Finally, Sec. VI summarises the results and presents conclusions.

\section{Non-Markovianity measures}

Let us begin by recalling basic definitions of the theory of open quantum systems.
The time evolution of the density matrix, describing the state of an open quantum system, is given by a $t$-parametrized family of completely positive and trace preserving (CPTP) maps $\Phi_t$, known as the dynamical map: $\rho_t = \Phi_t \rho_0$, with $\rho_0$ the density matrix of the open system at the initial time $t=0$. 
The dynamical map is divisible when it can be written as the composition of two CPTP maps $\Phi_{t} = \Phi_{t,t'} \Phi_{t',0} $, $ \forall t' \le t$. Non-divisibility therefore occurs if there exist times $t'$ at which $\Phi_{t,t'}$ is not CPTP.

A common feature of all non-Markovianity measures described in the following subsections is that they are based on the non-monotonic time evolution of certain quantities occurring when the divisibility property is violated. However, while non-monotonic behavior of such quantities always implies non-divisibility, the inverse is not true, i.e., there can be non-divisible maps consistent with monotonic dynamics. In this sense, if one would assume divisibility as the definition of non-Markovianity, all the other non-Markovianity measures should be considered as non-Markovianity witnesses.

\subsection{Rivas, Huelga, Plenio Measure }
With the defining attribute of all non-Markovian dynamics in mind, namely the violation of the divisibilty property, Rivas, Huelga and Plenio propose a measure based on the Choi-Jamiolkowski isomorphism \cite{Choi}. The inability to write the dynamical map $\Phi_{t}$ as a concatenation of two independent CPTP maps connects with intuitive reasoning of present dynamics being dependent on memory effects.  

For master equations written in the standard Lindblad form but with time-dependent coefficients,
\eqa
\frac{d \rho_t}{dt}&=& \mathcal{L}_t \rho_t= -i[H(t),\rho_t]\nn\\&+& \sum_k\! \! \gamma_k(t) \! \! \left(\! V_k(t) \rho_t V^{\dagger}_k(t)-\frac{1}{2}\{V_k^{\dagger}(t)V_k(t),\rho_t\}\!\right),
\eeqa
it is possible to show that the corresponding dynamical map satisfies divisibilty if and only if $\gamma_k(t)\geq0$. If on the other hand, $\gamma_k(t)$ becomes temporarily negative, there will exist an intermediate map $\Phi_{t,t'}$ which is not CPTP, defying the composition law. 

According to the Choi-Jamiolkowski isomorphism \cite{Choi}, $\Phi_{t,t'}$ is completely positive if and only if: 
\eq
(\Phi_{t,t'}\otimes \oper)\ket{\Omega}\bra{\Omega}\geq 0,
\eeq
where $\ket{\Omega}$ is a maximally entangled state of the open system with an ancilla. In light of this, one can quantify non-Markovianity by considering the departure of the intermediate map from a map which is completely positive \cite{RHP}: 
\eq
\mathcal{N}_{\text{RHP}}=\int_{0}^{\infty} dt\:g(t),
\label{RHP_One}
\eeq
where
\eq \label{g}
g(t)=\lim_{\epsilon\rightarrow0^+}\frac{||[\oper\otimes\oper+(\mathcal{L}_t\otimes\oper)\epsilon]\ket{\Omega}\bra{\Omega}||_{1}-1}{\epsilon}.
\eeq
and $\epsilon$ encapsulates the time elapsed between the initial time $t'$ to the final time $t$. The notation $||...||_1$ refers to the trace-norm.

Mathematically, an apparent advantage of this measure is that no optimization over states is required. The measure is most simple to calculate using only the specific form of the master equation. However, Eq. (\ref{g}) may be also calculated in terms of the intermediate dynamical map $\Phi_{t,t'}$  \cite{RHP}.  

We note here, in Ref. \cite{tony}, another measure based on criteria quantifying the deviation from Markovianity for quantum channels has been introduced.

\subsection{Breuer, Laine, Piilo Measure}
In Ref. \cite{BLP}, Breuer, Laine and Piilo introduce a measure based on the non-monotonicity of the trace distance in order to connect non-Markovian dynamics with a back-flow of information. The construction is based on the time evolution of the trace distance between two initial states, describing their relative distinguishability. The trace distance is contractive under CPTP maps therefore, for a divisible process, distinguishability of two initial states $\rho^{1,2}_0$ decreases continuously over time. Hence, the derivative of the trace distance, that can be seen as measuring the change of information content on the system, i.e. the information flux, is negative. In formulas 
\eq
\sigma(t,\rho^{1,2}_0)=\frac{d}{dt}D(\rho_t^1,\rho_t^2) \leq 0,
\eeq
where
\eq
D(\rho^1_t,\rho^2_t)=\frac{1}{2}\text{tr}|\rho^1_t-\rho^2_t |,
\eeq
is the trace distance between two states $\rho^1_t$ and $\rho^2_t$ evolving under the influence of a certain divisible dynamical map $\Phi_t:\rho_0\rightarrow\rho_t$. The authors of Ref. \cite{BLP} then define as non-Markovian  a process for which, for certain time intervals, $\sigma(t,\rho^{1,2}_0) > 0$, i.e., information flows back into the system.

Following this, the measure of non-Markovianity $\mathcal{N}_{\text{BLP}}$ is found by summing over all periods of non-monotonicity of the information flux, including an optimization over all pairs of initial states of the system: 
\eq
\mathcal{N}_{\text{BLP}}(\Phi_t)=\max_{\rho^{1,2}_0}\int_{\sigma>0}dt\:\sigma(t,\rho^{1,2}_0).
\eeq
Non-Markovian processes defined in this way are always non-divisible, however the converse is not necessarily true \cite{converse1}-\cite{converse3}.

The non-additivity of this measure has been numerically proven, highlighting the challenge presented when one wishes to consider higher dimensional systems of qubits  \cite{Brazil}-\cite{elsitwoqubit}. Proofs of specific mathematical attributes of the optimal state pairs have to some degree eased the numerical challenges of this calculation by elimating regions of the $n$-dimensional Hilbert space one should consider \cite{Optimal}. Indeed, it has been shown that the states which maximize the measure must lie on the boundary of the space of physical states and must be orthgonal. Finally, we note that in the spirit of this measure, trace distance is not a unique monotone distance and one may also use others such as the statistical distance \cite{MD}. 

\subsection{Luo, Fu, Song Measure } % REF \cite{NM_LFS}
Luo, Fu and Song in Ref. \cite{LFS} introduce a measure of non-Markovianity based on the monotonicity property characterizing the time-evolution of correlations between system and ancilla when the dynamics are divisible. They focus on the total correlations, including both classical and quantum, captured by the quantum mutual information 
\eq
I(\rho_{SA})=S(\rho_S)+S(\rho_A)-S(\rho_{SA}),
\eeq
where $\rho_S=\rm{tr}_A\rho_{SA}$ and $\rho_A =\rm{tr}_S \rho_{SA}$ are marginal states of a system and ancilla, respectively, and $S(\rho)$ is von Neumann entropy of state $\rho$. 
%This approach is supported by the monotonicity in time of $I(\rho_{SA})$ together with divisibility of dynamical map. 
The definition of this measure of non-Markovianity goes as follows
\eq
\mathcal{N}_{\text{LFS}}(\Phi)=\sup_{\rho^{SA}} \int_{\frac{d}{dt} I >0} \frac{d}{dt} I(\rho_t^{SA}) dt, \label{troppocomplicata}
\eeq
done over all possible initial states $\rho_{SA}$ with arbitrary Hilbert space of the ancilla and $\rho^{SA}_t=(\Phi_t \otimes \mathbb{I}) \rho_{SA}$. The authors insist that the optimization of the formula above is done over all possible initial states $\rho_{SA}$, where not only the state of the ancilla but also its Hilbert space is arbitrary. Of course this makes the optimization problem extremely complicated. 
Therefore the authors \cite{LFS} also propose a simpler version of this measure without optimization
\begin{equation}
\mathcal{N}_{\text{LFS}_0}(\Phi)= \int_{\frac{d}{dt} I >0} \frac{d}{dt} I(\rho_t^{SA}) dt, \label{tropposemplice}
\end{equation}
where $\rho^{SA}_t=(\Phi_t \otimes \mathbb{I}) |\Psi \rangle \langle \Psi|$ and $|\Psi \rangle$ is an arbitrary maximally entangled state.

Here we propose a slight simplification, a measure that is more general than Eq. (\ref{tropposemplice}) but simpler than Eq. (\ref{troppocomplicata}), and still has a significant interpretation. When the initial state $\rho_{SA}$ is pure, and also the initial state of the environment with which the system is interacting, one can rewrite $I(\rho_{SA})$ as the mutual information between the input and output of the channel defining the system evolution,
\eq\label{NI}
I(\rho, \Phi_t)=S(\rho)+S(\Phi_t \rho )-S(\rho,\Phi_t), 
\eeq
with $S(\rho)$ the von Neumann entropy of the input state, $S(\Phi_t \rho)$ the entropy of the output state and $S(\rho,\Phi_t)=S(\tilde \Phi_t \rho)$ the entropy exchange, i.e., the entropy at the output of the complementary channel $\tilde\Phi_t$. Now the non-Markovianity measure reads as
\eq
\mathcal{N}_{\text{I}}(\Phi)=\sup_{\rho} \int_{\frac{d}{dt} I >0} \frac{d}{dt} I(\rho,\Phi_t) dt.
\eeq

The main advantage of $\mathcal{N}_{\text{I}}$ over the original measure $\mathcal{N}_{\text{LFS}}$ is that the optimization has to be done only over the input state of the system. 

%For independent identical channels the measure $\mathcal{N}_\text{I}$ is at least additive, $\mathcal{N}_\text{I}(\Phi^{\otimes n}) \geq n \mathcal{N}_\text{I}(\Phi)$. In Ref. \cite{Brazil} the authors check this for 3 different models calculating the measure numerically for up to $5$ qubits. However, this is actually easy to prove analytically in general. Here we give the proof in Appendix {\color{red}[BOGNA: SAY HERE WHERE!]}.

\subsection{Entanglement-Assisted Classical Capacity Measure}
In Ref. \cite{Bogna}  two measures that link non-Markovian dynamics with an increase in the efficiency of quantum information processing and communication are introduced. This is motivated by observing that certain capacities of quantum channels are monotonically decreasing functions of time if the channel is divisible. This behavior is a consequence of the connection between the data processing inequality \cite{dpe} and the divisibility of the dynamical map. In other words, the measures proposed in \cite{Bogna} can be treated as witnesses of non-Markovian dynamics that cause revivals of quantum channel capacities. 

In that article the authors are concerned with two types of capacities. The first one is the entanglement-assisted classical capacity $C_{ea}$, which sets a bound on the amount of classical information that can be transmitted along a quantum channel when one allows Alice and Bob to share an unlimited amount of entanglement \cite{CQ}. It  is defined in terms of the quantum mutual information $I (\rho, \Phi_t)$ between the input and the output of the channel, as given by Eq. (\ref{NI}), by optimizing over all initial states $\rho$
\eq
C_{ea}(\Phi_t)=\sup_{\rho} I(\rho, \Phi_t). \label{eq:Cea}
\eeq

Since the entanglement-assisted capacity is monotonically decreasing in time when the channel is divisible, any increase of $C_{ea}$ would indicate violation of the divisibility property and so can be considered as a signature of non-Markovianity. Based on this a measure of non-Markovianity was introduced
\begin{equation}\label{NC}
\mathcal{N}_{\text{C}}=\int_{ \frac{d C_{ea}}{dt}(\Phi_t) >0} \frac{dC_{ea}(\Phi_t)}{dt} dt,
\end{equation}
where the integral is extended to all time intervals over which $dC_{ea}/dt$ is positive.

The measure presents some advantages with respect to, e.g., $\mathcal{N}_{\text{BLP}}$. The first advantage is that the calculation requires optimization only over the input state $\rho$, while the optimization required in the definition of $\mathcal{N}_{\text{BLP}}$  has to be done over pairs of initial states making it much more complicated to calculate, even knowing that we can restrict ourselves to only orthogonal pairs of states (having in mind that beyond the one qubit case more than one corresponding orthogonal state may exist). 

The second advantage is the additivity property. Thanks to the additivity of the mutual information one can prove the additivity of  $\mathcal{N}_\text{C}$ in the case of $n$ identical independent channels, i.e., $\mathcal{N}_\text{C}(\Phi^{\otimes n})=n \mathcal{N}_\text{C}(\Phi)$.

\subsection{Quantum Capacity Measure } 
The second quantity discussed in Ref. \cite{Bogna} is the quantum capacity $Q$. It gives the limit to the rate at which quantum information can be reliably sent down a quantum channel and is defined in terms of the coherent information between the input and output of the quantum channel $I_c (\rho, \Phi_t)$ \cite{CQ}:
\eq\label{eq:Q}
Q(\Phi_t) = \lim_{n \rightarrow \infty} \frac{\max_{\rho_n }I_c(\rho_n,\Phi_t ^{\otimes n})}{n},
\eeq
with $ I_c(\rho, \Phi_t)=S(\Phi_t \rho)-S(\rho,\Phi_t)$ \cite{I}.

Following the same line of reasoning done for $C_{ea}$,  a measure of non-Markovianity based on the non-monotonic behavior of the quantum capacity was introduced
\eq \label{NQ}
\mathcal{N}_\text{Q}=\int_{\frac{d Q(\Phi_t)}{dt} >0} \frac{d Q(\Phi_t)}{dt} dt,
\eeq

It it worth noting that the two measures of non-Markovianity based on capacities, in general, do not coincide even for degradable channels. The distinction between them is actually quite subtle. We notice indeed that as $ I(\rho,\Phi_t) = S(\rho)+I_c(\rho, \Phi_t)$, with $\rho$ the input state, we have $\frac{d}{dt} I(\rho,\Phi_t) = \frac{d}{dt} I_c(\rho, \Phi_t)$. Therefore a measure based on the violation of the data processing inequality for certain non divisible maps and $\mathcal{N}_\text{I}$ would not be able to distinguish between an increase in the different types of correlations. The optimizing state in the definitions (\ref{eq:Cea}) and (\ref{eq:Q}), however, is time dependent and does not coincide for the two quantities, hence  $dQ/dt \neq dC_{ea}/dt$.

Contrarily to the entanglement-assisted classical capacity, the quantum channel capacity is in general not additive. However, for degradable channels \cite{degrad}, the general definition coincides with the one-shot capacity, $Q(\Phi_t) = \max_{\rho }I_c(\rho,\Phi_t) $, and additivity for identical independent channels holds.

\section{Properties and interpretation of the non-Markovianity measures}

In this section we review some properties and physical interpretations of the considered non-Markovianity measures.

{\it 1) Physical interpretation}

Most of the non-Markovianity measures discussed here were born from an attempt to quantify and unveil the so-called reservoir memory effects. The manifestation of such memory effects, stemming from long lasting and non-negligible system-reservoir correlations, leads to a partial recovery of quantum properties previously lost due to the destructive effects of noise induced by the environment. 

The BLP measure, proposed in Ref. \cite{BLP}, was arguably the first attempt in this direction. The distinguishability between pairs of states, indeed, can be seen as a way to quantify the amount of information present in a quantum system. The more we know about the system, i.e., the more information we have, the more we are able to distinguish between quantum states. The environment's action of continuously monitoring the system gradually causes a loss of information and therefore a decrease of state distinguishability. In this description memory effects lead to information backflow, i.e. a temporary increase of distinguishability. 

It is worth noting here a common abuse of this interpretation. When talking of BLP non-Markovianity many authors refer to the flow of information from the system to the environment and then back into the system or, equivalently, of a back flow of information previously lost in the environment. The BLP definition, however, is based on quantities defined on the Hilbert space of the system only and in no way takes into account the information content of the environment or how it changes due to the interaction with the system.

In order to answer to this question, and therefore to look at a connection between changes of information in the system and in the environment, we started investigating  the capacity-based measures \cite{Bogna}. The entropy exchange term appearing in Eq. (\ref{NI}) is indeed the change in entropy of the environment, and the von Neumann entropy is one of the most common ways of quantifying the information content of a quantum state. A similar consideration holds for the LFS measure. However, while the LFS measure associates non-Markovianity to a memory-induced restoration of previously lost total (quantum + classical) correlations between an open quantum system and an ancilla, as measured by the mutual quantum information, the capacity-based measures look at a temporary increase of the entanglement assisted and quantum channel capacities. The latter ones, therefore, physically measure the total increase, due to reservoir memory, of the maximum rate at which information can be transferred in noisy channels for a fixed time interval or a fixed length of the transmission line.

The RHP measure, associated to divisibility, is often criticized for not having a clear physical interpretation and for being rather a mathematical definition than a physical one. Here we would like to conjecture, however, that a physical interpretation can be given in terms of the non-Markovian quantum jumps unravelling \cite{quantumjumps}. Whenever the system can be described in terms of a time-local master equation in Lindblad form with time-dependent coefficients, indeed, one can describe the dynamics of the open system in terms of an ensemble of state vectors whose evolution consists of non-Hermition deterministic dynamics interrupted by random quantum jumps with statistics connected to the time-dependent rates of the master equation. It is shown in Ref. \cite{quantumjumps} that when the rates become negative, i.e., the dynamical map is non-divisible and the RHP measure is non-zero, reverse quantum jumps restoring previously lost coherence occur. In this sense reverse jumps would be the physical manifestation of memory effects quantified by the RHP definition. Reverse quantum jumps always cancel or undo previously occurred jumps and therefore a jump-reverse jump pair describes a virtual process that is in principle not directly observable. 

A rigorous mathematical theory linking the RHP and the BLP measures has been presented in Ref. \cite{Darek}, while a general connection between the other measures presently does not exist and, we believe, is unlikely to be found. 

{\it 2) Experimental implementability}

Simulation of open quantum systems in both Markovian and non-Markovian regimes is nowadays in the grasp of the experimentalists \cite{trappedions1}-\cite{Matteo}. It is therefore important to identify the minimum requirements for implementing experiments measuring non-Markovianity. All non-Markovianity measures here studied cannot be written in terms of system's observables. Presently, all proposed witnesses of non-Markovianity rely on the full knowledge of the density matrix at all times $t$ \cite{WITN}. So, generally, experiments aimed at revealing one of the measures require quantum process tomography. But that is not all. The only measure that does not require an optimization procedure is the RHP measure that, however, assumes the knowledge of either the dynamical map or the explicit form of the master equation. In contrast, the BLP, LFS, and BCM measures do not in principle require an assumption on the specific model of open quantum system dynamics. However this comes with the heavy overload that the optimization should be performed experimentally. 
Alternatively, one may assume the validity of a given model and solve the optimization problem either numerically or analytically. In this case it is generally sufficient to know the evolution of the density matrix, so once again process tomography is generally required. From this more realistic perspective the requirement for experimental implementation of all measurements are comparable. We conclude by noticing that both the BLP and the RHP measures have been experimentally observed in optical experiments simulating pure dephasing environments \cite{puremodel1}-\cite{puremodel3}.

{\it 3) Interest for quantum technologies}

We conclude this section with a few remarks on the potential usefulness of non-Markovianity for quantum technologies. Recent results have shown that, in certain circumstances, the manipulation of reservoir spectral properties may lead to improvements in certain quantum technologies \cite{MNonM1}-\cite{MNonM6}. First of all, it is worth noticing that the improvements demonstrated for quantum metrology \cite{MNonM1} and for quantum key distribution \cite{MNonM2} do hold also when the dynamical map is divisible, i.e., they are not directly connected to memory effects, as they may occur also when the dynamics is Markovian according to the measures here discussed. Other results work instead in the specific case of correlated dephasing environments, where the effect of the noise is known, such as the quantum teleportation scheme of Ref. \cite{MNonM3}. The only general answer to the question of the connection between certain quantum technologies and non-Markovian memory effects is given in Ref. \cite{Bogna} and was in fact one of the motivations of the introduction of the channel capacity measures. Whether or not non-Markovianity can be seen as a resource for quantum technologies is still an open problem and would require rephrasing the question in terms of resource theory. We hope that such a direction will be the object of future studies.

%
%We conclude the section by noting that, as there is no general monotonicity relation between the different non-Markovianity measures, it does not make sense to compare their absolute values. Therefore, in the paper we renormalize all measures to take values between zero and 1, and we look at both their qualitative behavior and the Markovian to non-Markovian crossover when certain physical parameters of the model are changed.

\section{Pure Dephasing Dynamics}
We begin our analysis by looking at the dynamics of one or two qubits interacting with environments leading to pure dephasing. Note that, as there is no general monotonicity relation between the different non-Markovianity measures, it does not make sense to compare their absolute values. Therefore, in the paper we renormalize all measures to take values between zero and 1, and we look at both their qualitative behavior and the Markovian to non-Markovian crossover when certain physical parameters of the model are changed.

A microscopic model of the total system-environment dynamics is presented in Refs. \cite{puremodel1}-\cite{puremodel3}. One of the advantages of this model is that it is amenable to an exact solution \cite{puremodel1}-\cite{puremodel3}. We will consider initially the case of a single qubit interacting with a reservoir with spectral density of the Ohmic class. The behavior of all non-Markovianity measures in this case has been studied in Refs. \cite{RHP}-\cite{BLP}, \cite{NMPure3}-\cite{NMPure4}. We will then extend the analysis to the case of two qubits in both independent and common environments. Both BLP and RHP measures have been studied also in this case in Ref. \cite{Brazil}-\cite{elsitwoqubit}. The analysis of the LFS and BCM measures for two qubits is the first result of our paper.

In the following subsection we will present results on the behavior of the four non-Markovianity measures here considered. A thorough comparison and discussion about these results will be given in Sec. VI.

\subsection{Single qubit: the model}
The dynamics of a purely dephasing single qubit is captured by the time-local master equation \cite{bp}:
\eq \label{twoqubit}
\mathcal{L}_t\rho_t=\gamma_1(t)\left[\sz\rho_t\sz-\rho_t\right],
\eeq
with $\gamma_1(t)$ the time-dependent dephasing rate and $\sz$ the Pauli spin operator. The decay of the off-diagonal elements of the density matrix is described by the decoherence factor $e^{-\Gamma(t)}$, where $\Gamma(t)\geq 0$ and, for zero-temperature environments \cite{puremodel2},
\eqa
\Gamma(t)&=&2\int_{0}^{t} dt' \: \gamma_1(t')\nn\\&=&4\int d\omega \: J(\omega) \frac{1-\cos(\omega t)}{\omega^2},
\label{gamma0}
 \eeqa
with $J(\omega)$ the reservoir spectral density \cite{puremodel2,NMPure4}.

We consider a reservoir spectral density of the form,
 \eq
 J(\omega)=\frac{\omega^s}{\omega_c^{s-1}} e^{-\omega/\omega_c}, 
 \eeq
where $\omega_c$ is the cutoff frequency and $s$ is the Ohmicity parameter. Further details on this model can be found in Appendix A. 

In this model re-coherence occurs when $\Gamma(t)$ temporarily decreases for certain time intervals, corresponding to a negative value of the dephasing rate $\gamma_1(t)$. 
One may analytically determine the times $t\in[a_i,b_i]$ encapsulating non-monotonic intervals of $\Gamma(t)$, i.e. corresponding to $\gamma_1(t)=0$, with $i=1, 2, 3,...$ the number of such time intervals. The extremes of the time intervals, $a_i$ and $b_i$, will depend on the Ohmicity parameter $s$, as changing $s$ one changes the form of the reservoir spectral density.
In Ref. \cite{Timediscord} it is shown that, for $s\leq2$,  $\gamma_1(t) > 0$ at all times, or equivalently $\Gamma(t)$ increases monotonically. For $2<s\leq 4: a_1=\tan\pi/s, b_1=\infty$ and for  $4<s\leq 6: a_1=\tan\pi/s, b_1=\tan2\pi/s$, i.e., we only have one time interval of non-monotonic behavior. For $s>6$, $i>1$, i.e., there are more than one interval of time for which the dephasing rates become negative. 

We give here the analytical expressions of $\Gamma(t)$ at times $a_i$ and $b_i$ as these will be used in the following. 
\eqa
& & \Gamma(a_1)=\frac{2\tilde\Gamma[s][1+\cos^{s}(\pi/s)]}{s-1} \label{1gamma1} \\ & & 
\Gamma(b_1 )=\frac{2\tilde\Gamma[s][1-\cos^{s}(2\pi/s)]}{s-1} \hspace{0.7 cm} 4<s\leq 6 \\ &   & 
\Gamma(b_1)=2\tilde\Gamma[s-1] \hspace{2.6 cm} 2<s\leq 4, \label{1gammb1}
\eeqa
where $\tilde\Gamma[x]$ is the Euler gamma function

\subsection{Single qubit: the measures}
\begin{figure}
\includegraphics[width=0.36\textwidth]{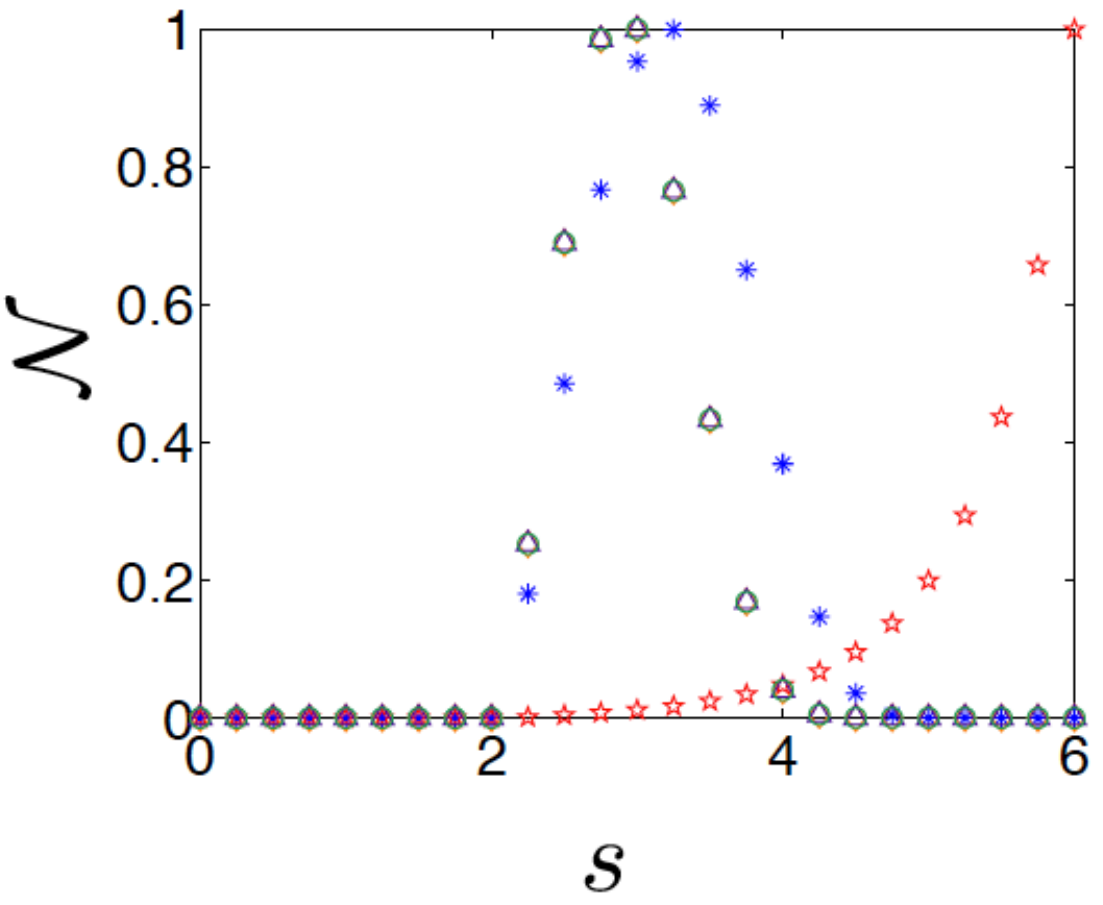} 
\caption{(Color Online) Non-Markovianity measures for a single purely dephasing qubit as a function of the Ohmicity parameter $s$. We show the RHP Measure (red star),  the BLP (blue asterisk), the LFS measure (green circle), the quantum capacity measure (purple triangle) and the entanglement-assisted capacity measure (orange diamond). Note that the last three measures in this case coincide. All the measures in this case are normalized to unity. Note that the value of all measures for $s>2$ is always non-zero, even if all measures except RHP take very small values for $s>4$.}
\label{OQ}
\end{figure} 
{\it RHP Measure}

Inserting Eq. (\ref{twoqubit}) into Eq. (\ref{g}) and using Eq. (\ref{RHP_One}) one immediately obtains the analytical expression for the RHP non-Markovianity measure $\mathcal{N}_{\text{RHP}}$ for a single qubit \cite{RHP}:
  \eq
\mathcal{N}_{\text{RHP}}=-2\int_{\gamma_1(t)<0}dt\:\gamma_1(t)=\sum_i\Gamma(a_i)-\Gamma(b_i). \label{RHP dephasing}
  \eeq
For the sake of simplicity we look at values of the Ohmicity parameter in the interval $0 \le s \le 6$. In this case there is only one interval of negativity of the decay rates and the only values needed are $\Gamma(a_1)$ and $\Gamma(b_{1})$, defined in Eqs. (\ref{1gamma1})-(\ref{1gammb1}).

%For zero temperature, it is straightforward to prove the dephasing rate $\gamma_1(t)$ takes temporarily negative values if and only if $s>s_{\text{crit}}=2$ [REF]. Hence, for super-ohmic environments $s>2$ only, the dynamics are non-divisible and $\mathcal{N}_{\text{RHP}}\neq 0$. 
In Fig. \ref{OQ} we plot $\mathcal{N}_{\text{RHP}}$ for different values of the Ohmicity parameter $s$ (red stars). As one can see from the analytical expression, for increasing $s$ the area of the region of negativity of the dephasing rates increases, hence the measure monotonically increases for higher and higher values of $s$.

{\it BLP Measure} 

It is straightforward to show that $\mathcal{N}_{\text{BLP}}$  can be written in terms of the two independent elements of the single qubit density matrix \cite{EM}
\eq
\mathcal{N}_{\text{BLP}}=-2 \max_{m,n} \int_{\gamma_1<0}dt\:\gamma_1(t)\frac{|n|^2e^{-2\Gamma(t)}}{\sqrt{m^2+|n|^2e^{-2\Gamma(t)}}}, 
\label{BLPone}
\eeq
where $m=\rho_{11}^1(0)-\rho_{11}^2(0)$ and $n=\rho_{12}^1(0)-\rho_{12}^2(0)$, with $\rho^i_{jj} (0)$  the diagonal elements of the initial density matrices  of the pair and $\rho_{jk}^i(0)$ their off diagonal elements, with $(i,j,k=1,2)$.  This expression shows immediately that  $\sigma(t)>0$ iff $\gamma_1(t)<0$; i.e. $\mathcal{N}_{\text{BLP}}\neq 0$ only when the dynamical map is non-divisible \cite{BLP}. 

For the model here considered it is possible to analytically solve the optimization problem of Eq. (\ref{BLPone}) \cite{ana}. The pair of states optimizing the increase of the trace distance are antipodal states lying on the equatorial plane, e.g., the states $\ket{\pm}=\frac{1}{\sqrt{2}}(\ket{0}\pm\ket{1})$, with $\ket{0}$ and $\ket{1}$ the two states forming the qubit.
%
%Optimizing over antipodal states on the boundary of the Bloch sphere numerically proves the maximizing initial state pair is $\ket{\pm}\bra{\pm}$. The basis we use to define these states refer to a superposition of the ground and excited state of the system, $\ket{\pm}=\frac{1}{\sqrt{2}}(\ket{0}\pm\ket{1}$. 
Hence, we now have:
\eq
\mathcal{N}_{\text{BLP}}=\sum_i e^{-\Gamma(b_i)}-e^{-\Gamma(a_i)} \label{NBLPdephasing}
\eeq
where again $t\in[a_i,b_i]$ indicates the time intervals when $\gamma_1(t)<0$.

Figure \ref{OQ} shows the behavior of $\mathcal{N}_{\text{BLP}}$ when changing $s$ (blue asterix). The measure remains non-zero for increasing values of $s$ but, contrarily to the RHP measure it starts decreasing taking small but finite values for $s>3.2$. 

{\it LFS Measure} 

Numerical results show that the optimizing state for this measure is the maximally mixed state \cite{Bogna}. Since 
$I\left(\frac{\mathbb{I}}{2}, \Phi_t \right)= 2-H_2 \left( \frac{1}{2}+ \frac{e^{-\Gamma(t)}}{2}\right)$, where $H_2(.)$ stands for the binary Shannon entropy, one has
\begin{equation}
    \frac{d}{dt} I\left(\frac{\mathbb{I}}{2}, \Phi_t \right)=-\frac{1}{2} \gamma_1(t) e^{-\Gamma(t)} \log_2 \left(\frac{1+ e^{-\Gamma(t)}}{1- e^{-\Gamma(t)}}\right),
\end{equation}
which indicates that the measure $\mathcal{N}_{\text{I}}$ has non-zero value if and only if $\gamma_1(t)<0$, i.e., whenever the dynamical map is non-divisible.

The explicit expression for the measure $\mathcal{N}_{\text{I}}$ is then easily written as 
 \begin{equation}\label{NID}
\mathcal{N}_\text{I}(\Phi_t )=\sum_{i} \left[H_2 \left( \frac{1}{2}+ \frac{e^{-\Gamma(a_i)}}{2}\right)-H_2 \left( \frac{1}{2}+ \frac{e^{-\Gamma(b_i)}}{2}\right) \right].
\end{equation}
%where $t\in[a_i,b_i]$ are the time intervals for which $\gamma_1(t)<0$ specified in Sec. III A.

It is worth noting that in the case of the dephasing channel here considered, for the simple measure without optimization we have, $\mathcal{N}_\text{I}(\Phi_t)=\mathcal{N}_{\text{LFS}_0}(\Phi_t)$.

In Fig. \ref{OQ} we plot $\mathcal{N}_{\text{LFS}}$ as a function of $s$ (green circles). We note that the behavior of this measure is qualitatively similar to that of $\mathcal{N}_{\text{BLP}}$.

{\it BCM Measures } 

In case of pure-dephasing channels the state optimizing the formula for the classical entanglement assisted capacity is a maximally mixed state, independently of time or of the specific properties of the environmental spectrum. This means that there is a simple analytical formula characterizing it, namely  $C_{ea}^D=I(\frac{\mathbb{I}}{2}, \Phi_t)$ \cite{QCDegradable}. Hence the measures based on mutual information and classical entanglement assisted capacity, in this particular case, coincide: $\mathcal{N}_{\text{C}}(\Phi_t)=\mathcal N_\text{I}(\Phi_t)$. 

The dephasing channel is degradable for all admissible dephasing rates $\gamma(t)$, i.e. whenever $\Gamma(t) \ge 0$. This simplifies the calculations of the quantum capacity. Indeed, we find that the state optimizing the coherent information in the definition of the quantum capacity is once again the maximally mixed state. Having this in mind one can show a very simple relation between the two capacities, namely $C_{ea}^D(t) = 1+ Q^D(t)$. It follows immediately that $\mathcal{N}_\text{Q}(\Phi_t)=\mathcal{N}_\text{C}(\Phi_t)=\mathcal N_I(\Phi_t)$. \\

We conclude this subsection stressing that, for the single qubit case, all non-Markovianity measures detect non-divisibility, hence when studying their behavior as a function of the parameter $s$ we obtain the same crossover between Markovian and non-Markovian dynamics, i.e., $s=2$. A direct comparison between the analytic expressions of Eqs. (\ref{RHP dephasing}), (\ref{NBLPdephasing}) and (\ref{NID}) clarifies the different qualitative behaviour shown by the RHP measure with respect to the other ones in Fig. 1. Indeed, the increasing value of the former measure is due to the fact that, for increasing $s$, the number of periods of negativity of the dephasing rate increases and with it the terms contributing to the sum of  Eq. (\ref{RHP dephasing}). On the contrary, the other measures all depend on the dephasing factors $e^{-\Gamma(t)}$ calculated at times at which the direction of information flow changes, i.e. $t=a_i$ and $t=b_i$. For increasing values of $s$, in the $s \gtrsim 3$ parameter space, however,  $e^{-\Gamma(a_i)} \simeq e^{-\Gamma(b_i)}$, hence the values of both $\mathcal{N}_\text{BLP}$ and $\mathcal{N}_\text{I}(\Phi_t)=\mathcal{N}_\text{Q}(\Phi_t)=\mathcal{N}_\text{C}(\Phi_t)$ decreases, as one can easily see from Eqs. (\ref{NBLPdephasing}) and (\ref{NID}). 

From a physical point of view, having in mind the interpretations that we discussed in Sec. III, this means that, while the number of reverse jumps always increases with $s$ (as the number of periods of negativity in the dephasing rate increases), the information backflow has a maximum for a certain values of $s \simeq 3$ and then decreases due to the fact that the amplitude of oscillations in the decay rates become smaller and smaller. 

\subsection{Two qubits in independent environment: the model}

We now turn our attention to a bipartite system consisting of two qubits, $A$ and $B$, individually coupled to their own identical and non-correlated environment. The dynamical map in this case is given by  $\Phi_t^{AB}=\Phi_t^A\otimes\Phi_t^B$, and the corresponding master equation is the sum of two identical Lindblad terms of the form of Eq. (\ref{twoqubit}), describing the dynamics of each qubit, both characterized by the same dephasing rate $\gamma_1(t)$.

{\it RHP Measure }  

From the form of the master equation one sees immediately that the RHP measure for two independent qubits is exactly additive. The measure has a simple expression for any form of spectral density, i.e. of dephasing rate $\gamma_1(t)$.
%  \eq
%\mathcal{N}_{\text{RHP}}=-4\int_{\gamma_1(t)<0}dt\:\gamma_1(t)=-2\sum_i\Gamma(b_i)-\Gamma(a_i).
%  \label{independent}
%  \eeq
For the Ohmic class of spectra we can again plot $\mathcal{N}_{\text{RHP}}$ as a function of $s$ obviously obtaining exactly the same qualitative behavior as the one of Fig. \ref{OQ}.

{\it BLP Measure}

Already for the most straightforward generalization of the single qubit channel, namely two qubits in identical independent environments, the problem of finding the optimal pair of states maximizing the increase of the trace distance is non trivial. Even more because in practice we need to fix the time interval over which we describe the evolution and, in general, the optimal pair does depend on the chosen time interval. Here we have performed extensive numerical optimization using random pairs of states (See also the plots in Appendix 2). We have compelling evidence that the optimal states in this case are $\ket{\pm\pm}\bra{\pm\pm}$, i.e., product states of the single qubit optimal pairs, as shown in Fig \ref{A1}. 

We note in passing that these optimal states do depend on the specific form the the spectrum. For the Ohmic class here considered they are always of the form $\ket{\pm\pm}\bra{\pm\pm}$, but for other forms of spectral densities such as the one of the Bose-Einstein condensate reservoir \cite{Bose} of Ref. \cite{Carole} the optimal pair is different (pair of Bell states), despite the fact that the operatorial form of the master equation is the same of the one here considered.

Having this in mind, and remembering the form of the dynamical map $\Phi^{AB}_t$, one obtains that for this specific model the BLP measure for two independent qubits is exactly that of one qubit, i.e., it is given by Eq. (\ref{NBLPdephasing}).
%\eq
%\mathcal{N}_{\text{BLP}}=\sum_i e^{-\Gamma(b_i)}-e^{-\Gamma(a_i)},
%\eeq
%where $t\in[\textcolor{red}{\tilde a_i,\tilde b_i]}$ indicates the time intervals when $\gamma_1(t)<0$. 
%%Nonetheless, non-apparent change to the information flux as a result of an additional qubit is physically counterintuitive and contrary to what one would expect. 
We stress once again that this result relies on the specific form of the decoherence factors that we have calculated for the Ohmic class of reservoir spectral densities. In the model of Ref. \cite{Carole}, on the contrary, the measure is sub-additive.

{\it LFS Measure}
 
Numerical evidence indicates that the optimizing state for the measure $\mathcal{N}_\text{I}$ is the two-qubit maximally mixed state. Notice that this is a product state of states optimizing the single qubit channel discussed in Sec. IV A, hence the additivity property is satisfied $\mathcal{N}_\text{I}(\Phi^{AB}_t)=2\mathcal{N}_I(\Phi^{A}_t).$ As in the single qubit case here again the simplified measure $\mathcal{N}_{\text{LFS}_0}$ gives still the exact value of $\mathcal{N}_{I}$.

{\it BCM Measures} 

For any number $N$ of qubits interacting with independent identical environments the dephasing channel is degradable for all values of parameters. Hence both measures based on capacities of quantum channels are additive. Moreover, having in mind that in the case of one-qubit dephasing channel both capacity measures are equal to $\mathcal{N}_\text{I}$, which is also additive, we can conclude that again these three measures are equivalent, $\mathcal{N}_\text{Q}(\Phi^{AB}_t)$=$\mathcal{N}_\text{C}(\Phi^{AB}_t)=\mathcal{N}_\text{I}(\Phi^{AB}_t)$. 

We conclude this subsection by noticing that, when compared to the single qubit case, the two-qubits in independent environments presents no new features. A comparison of the renormalized measures gives a figure that is exactly identical to Fig. \ref{OQ}. All measures in this case are additive, except for the BLP measure having the property that the measure for one qubit is the same as the measure for two qubits. 

\begin{figure*}[htp]
\centering
\includegraphics[width=0.9\textwidth]{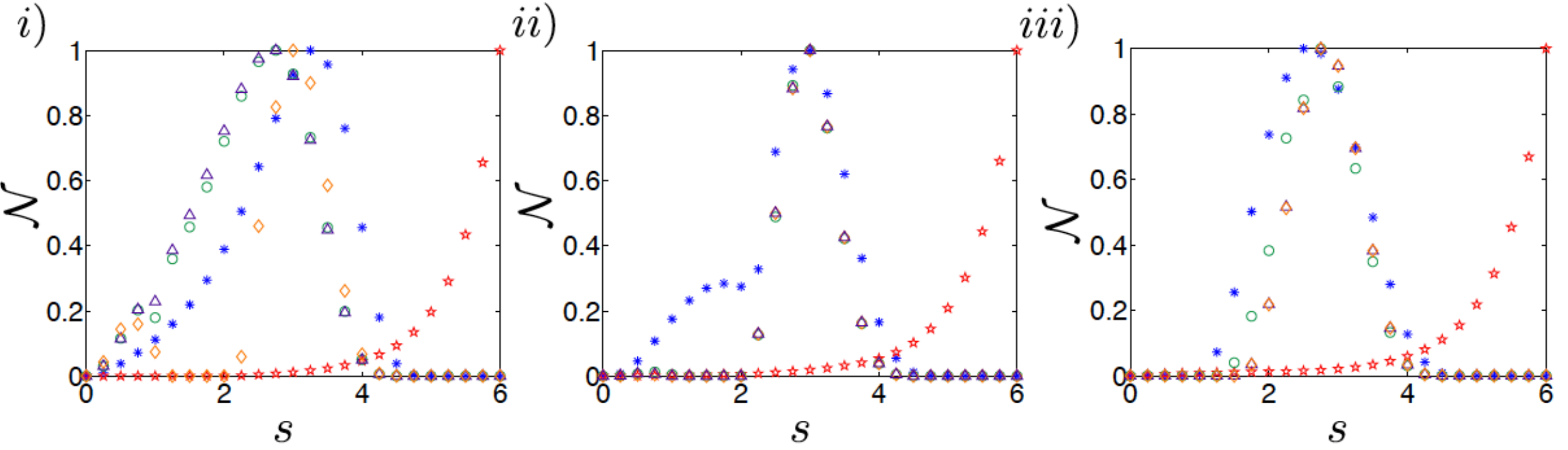}
\caption{(Color Online) Non-Markovianity measures for two-qubit systems interacting with a common pure dephasing environments for transit times $t_s$ i) 0.25, ii) 2 and iii) 6. We plot the RHP measure (red star),  the BLP measure (blue asterisk), the LFS  measure (green circle), the quantum capacity measure (purple triangle) and the entanglement-assisted capacity measure (orange diamond). All the measures in this case are normalized to unity and plotted against the Ohmic parameter $s$.}
\label{TQ}
\end{figure*}

\subsection{Two qubits in common environment: the model}

The next level of generalization of the simple dephasing model consists in assuming that the two qubits see a common environment inducing pure dephasing. We will consider specifically the model described in Ref. \cite{puremodel2}. As far as we know this is the first time that for this model the measures here considered are investigated and compared. One of the properties of common environments is the presence of a cross-talk term acting as an effective reservoir-mediated interaction between the qubits. Intuitively we expect that the presence of this term may affect additivity property present in the individual environments case.

The master equation describing the dynamics of the composite system is given in Refs. \cite{Carole,ME}. The time evolution is now described by two time-dependent coefficients $\gamma_1(t)$ and $\gamma_2(t)$. The former one is the dephasing rate of the single qubit case appearing in Eq. (\ref{twoqubit}) while the latter one is the cross-talk term mentioned above. The master equation is in Lindblad form with time dependent dephasing rates $\frac{1}{2}\gamma_{\pm}(t)= \frac{1}{2}(\gamma_1(t) \pm \gamma_2 (t))$. 

Also in this case an exact analytic solution can be found. The density matrix at time $t$ takes the form  \cite{puremodel2},
\eqa
\rho_t =
 \begin{pmatrix}
 1&e^{-\Gamma(t)}&e^{-\Gamma(t)}&e^{-\Gamma_-(t)}\\
 \\
  e^{-\Gamma(t)}&1&e^{-\Gamma_+(t)}&e^{-\Gamma(t)}\\
  \\
  e^{-\Gamma(t)}&e^{-\Gamma_+(t)}&1&e^{-\Gamma(t)}\\
  \\
  e^{-\Gamma_-(t)}&e^{-\Gamma(t)}&e^{-\Gamma(t)}&1\\
 \end{pmatrix}\circ\rho(0), \:\:\:\:\:\:\:\:
\eeqa
where $\circ$ is the Hadamard product and,
 \eqa
 \Gamma_\pm(t)&=&2\Gamma(t)\pm\delta(t)\nn\\&=&8\int d\omega \: J(\omega) \frac{1-\cos(\omega t)}{\omega^2}\left(1\pm\cos\omega t_s\right), \nn\\ & & 
\label{gammapm}
\eeqa
with $\Gamma(t)$ given by Eq. (\ref{gamma0}) and 
\eq
\delta(t)=4\int_0^t dt'\:\gamma_2(t').
\label{delta}
\eeq
The transit time $t_s$ describes the time it takes for a wave propagating at the characteristic speed of sound to travel from one qubit to the other, given that the qubit distance is ${R}$ \cite{puremodel2}. As before, for the sake of simplicity, we consider only the zero temperature reservoir case. 
%
%The corresponding Master equation is as for two independent qubits but with two additional Lindblad structured mixed terms, i.e., with operators corresponding to both qubits and time-dependent coefficient $\gamma_2(t)$ [REF]. These terms fully describe the collective dynamics, encapsulated by the so-called ``cross-talk" term \cite{Carole, Bose}:
%\eq
%\delta(t)=4\int_o^t ds\:\gamma_2(s).
%\label{delta}
%\eeq

%\subsubsection{Comparison between the measures}
{\it RHP Measure } 

From the form of the master equation it is straightforward to show that the RHP measure takes the form
    \eq
 \mathcal{N}_{\text{RHP}}=-2\int_{\gamma_+(t)<0}dt\:\gamma_+(t)-2\int_{\gamma_-(t)<0}dt\:\gamma_-(t).
  \label{RHP_Two}
  \eeq
Notice that it is sufficient that one of the coefficients appearing in the master equation is non-zero to violate divisibility and give a non-zero value of the measure.
If we now specify our analysis to the Ohmic class of spectral densities characterized by the parameter $s$ we can numerically prove that the measure is super-additive for any value of $s$.

In Fig. \ref{TQ} we plot $\mathcal{N}_{\text{RHP}}$ versus $s$ (red stars) for increasing values of separation between the qubits. Even if not clearly visible from the plots, there exists a critical value of $s$ in correspondence of which the dynamics changes from Markovian to non-Markovian. This value, however, depends on the distance between the qubits. We have calculated numerically that the critical value $s_c$ is $ \simeq 6.5 \times 10^{-2}$ for case i), $ \simeq 6.6  \times 10^{-2}$ for case ii) and $ \simeq 6.9  \times 10^{-2}$ for case iii) for the time period we consider. Generally $s_c$ increases for increasing values of the distance between the qubits, attaining its maximum value $s_c=2$ for $R \rightarrow \infty$ when $\gamma_2(t)\rightarrow 0$ and we re-obtain the case of independent environments. Therefore, for $R\rightarrow \infty$, the measure is additive in the sense that the measure of the two qubit case is equal to twice the measure for each individual qubit.

{\it BLP Measure }

The common dephasing environment shows in an exemplary way the subtle aspects connected to the optimization procedure in the definition of the BLP measure. In this case we have shown numerically that the optimizing pair depends both on the value of $s$ and on the distance ${R}$ between the qubits, as both are important parameters in the effective spectrum seen by the qubits. More precisely, the changes in the optimizing pair stem from the  complex evolution of the cross-talk term $\delta(t)$.

Extensive optimization procedures allows us to conclude that the maximizing states are either the super-  or the sub-decoherent Bell states $\vert \Psi_{\pm} \rangle =\frac{1}{\sqrt{2}}(\ket{01}\pm\ket{10})$ and $\vert \Phi_{\pm} \rangle=\frac{1}{\sqrt{2}}(\ket{00}\pm\ket{11})$, depending on $s$ and ${R}$, for any finite value of ${R}$. This is shown in detail in Fig. \ref{A2} and discussed in Appendix 2. Note that the decoherence factors of the super- and sub-decoherent states are $\Gamma_+(t)$ and $\Gamma_-(t)$, respectively. 

The analytical expression for the BLP measure can be written as \cite{Carole}
\eq \label{analyticalBLP}
\mathcal{N}_{\text{BLP}}=\max\{\mathcal{N}_\Phi,\mathcal{N}_\Psi\},
\eeq
where $\mathcal{N}_\Phi=\sum_{i}e^{-\Gamma_-(b_i)}-e^{-\Gamma_-(a_i)}$ is the measure if the sub-decoherent Bell states form the maximizing pair and $\mathcal{N}_\Psi=\sum_{i}e^{-\Gamma_+(b_i)}-e^{-\Gamma_+(a_i)}$ is the measure if the super-decoherent Bell states form the maximizing pair. Time intervals $t\in[a_i, b_i]$ again indicate the periods of information backflow, manifested as $d\Gamma_\pm(t)/dt\sim\gamma_1(t)\pm\gamma_2(t)<0$.

From the analytic expression of $\mathcal{N}_{\text{BLP}}$ one sees immediately that, also in this case, BLP non-Markovianity coincides with non-divisibility. However, the qualitative behavior of the BLP measure when changing the reservoir spectrum is different from that of $\mathcal{N}_{\text{RHP}}$, as shown in Fig. \ref{TQ}. We note that, for a given value of $t_s$ (or equivalently of ${R}$), the change in the optimizing pair is clearly visible (see Figs. \ref{TQ} ii) and \ref{A2} a) ii)). 

Numerical investigation also shows that the measure is super-additive as a result of the qubit dephasing collectively through environment-mediated interactions, in contrast to the independent-environment case. For $R \rightarrow \infty$ the measure reverts to the independent case with optimal pairs $\ket{\pm\pm}\bra{\pm\pm}$ and super-additivity is lost. 

%${\it LFS Measure} \textcolor{red}{To be added}

{\it BCM and LFS Measures } %\textcolor{red}{To be added}

The effect of the cross-talk term in the considered model of common environment can be clearly observed  for all three measures, $\N_{\text{LFS}}$, $\N_\text{Q}$ and $\N_\text{C}$, especially when one compares the case when two qubits are very close to each other, $t_s=0.25$, with the two cases when we take the qubits further and further apart, for $t_s=2$ and $t_s=6$, see Fig. \ref{TQ}. 

Generally, the two channel capacity measures and the LFS measures have a similar behavior with the exception of the case  $t_s=0.25$ in which $\mathcal{N}_\text{C}$ presents a different feature when $s$ is varied. More specifically,  $\mathcal{N}_\text{C}$ has two peaks, a big one for $s\simeq 3$, and an additional small one for $s\simeq 0.75$. As the distance between the qubits increases, the small peak amplitude decreases and eventually vanishes, while the bigger peak moves towards the value typical of the independent environments case, i.e. $s=2.75$. 

This difference can be understood by looking at the optimizing states. Numerical investigation shows that when the qubits are closeby (Fig. 2 i)) the optimizing states of all three measures are of rank 2 with eigenvalues $\lambda_{1,2}=0.5\pm \epsilon$, where $\epsilon \in [ 0, 0.1)$. The corresponding values of $\N_{\text{LFS}}$ and $\N_\text{Q}$ differ from those that would be obtained had the optimal state been the maximally mixed one. On the contrary, due to the presence of the term $S(\rho)$ in the definition of the entanglement assisted capacity, $\mathcal{N}_{\text C}$ takes a value similar to the one it would have if the optimizing state were the maximally mixed one.

%For measure $\N_{LFS}$ and $\N_Q$ the value obtained for maximally mixed state is significantly smaller than the one obtained for the optimizing state, while for $\mathcal{N}_C$ this discrepancy is very small and eventually for values of $s >2$, the optimizing state becomes the maximally mixed state, as in the independent case. This happens due to the term $S(\rho)$ in expression (\ref{eq:Cea}), that is missing in definition of $Q$ (\ref{eq:Q}), which obviously takes the grates value for a maximally mixed state. 

When $t_s=2,6$  the optimizing state for both BCM measures is the maximally mixed state for all values of $s$. Notice that this is the same state that is optimal in case of independent environments and also here we obtain $\N_{\text C}=\N_\text{Q}$. Finally, for $\mathcal{N}_{\text{LFS}}$ the optimizing states are close to maximally mixed state, these are states of rank 4 with eigenvalues given by $0.25\pm \epsilon_i$, where $\epsilon_i \in[0,0.1)$ for $i=1,...,4$. 
From numerical analysis we also know that similarly to $\mathcal{N}_{\text{BLP}}$, the measures $\N_{\text{LFS}}$, $\N_{\text C}$ and $\N_{\text Q}$ are super-additive.

\begin{figure*}
\includegraphics[width=0.68\textwidth]{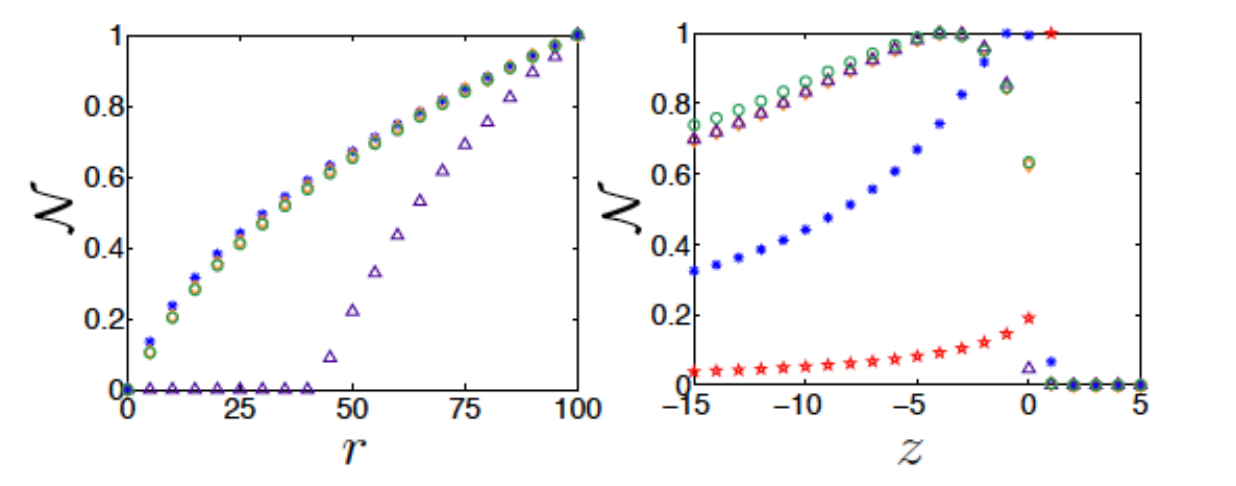} 
\caption{(Color Online) Non-Markovianity measures for a qubit undergoing amplitude damping for i) Lorentzian spectrum and ii) Photonic Band Gap model. We plot the RHP measure (red star),  the BLP measure (blue asterisk), the LFS measure (green circle), the quantum capacity measure (purple triangle) and the entanglement-assisted capacity measure (orange diamond). All the measures in this case are normalized to unity. We consider the following times periods: i) $0 \le \lambda t \le 40$ and ii)  $0 \le \beta t \le 20$. For the Lorentzian spectrum, the RHP measure is zero for $r \le 0.5$ while it diverges for $r>0.5$.}
\label{AD}
\end{figure*}

\section{Amplitude Damping Channel}
Let us now consider the case in which the interaction between the quantum system and its environment leads to energy exchange between the two, resulting in dissipative open system dynamics. As we did in Sec. IV we will focus on exemplary open system models amenable to an exact analytical solution as this allows us to gain a solid understanding of the physical phenomena associated with reservoir memory. 

Once again we proceed for increasing levels of complexity. We consider first the single qubit case interacting with a quantized bosonic field with both Lorentzian and Photonic Band Gap (PBG) spectra. For the single qubit Lorentzian case both the BLP and the BCM measures have been studied numerically in Refs. \cite{BLP} and \cite{Bogna}, respectively. While only the BCM measure has been  investigated before in the Photonic Band Gap model here used, the other measures indeed have not. We then discuss for the first time the generalisation to the case of two qubits immersed in two independent identical environments.

The common environment scenario is not considered here because both the LSF and the  quantum capacity measures present a high level of difficulty in this case. It seems indeed that the problem of calculating classical or quantum capacity for two qubits interacting with a common environment has never been considered in the literature. Here the optimization problem is amplified by the fact that it should be performed at each time instant of the evolution. 

\subsection{Single qubit: the model}
The dynamics of a single amplitude damped qubit is captured by the time-local master equation \cite{bp}: 
\eq 
\frac{d\rho_t}{dt}=\gamma_1(t)\left[\sigma_{-}\rho_t\sigma_{+}-\frac{1}{2}\{\sigma_{+}\sigma_{-},\rho_t\}\right],
\eeq
where $\sigma_{\pm}$ are the spin lowering and rising operators and
\eq
\gamma_1(t)=-2\Re\frac{\dot{G}(t)}{G(t)}. \label{gammaADC}
\eeq
The function $G(t)$ depends on the form of the reservoir spectral density and is given in Appendix A for the two models here considered.

The state of the density matrix of the qubit at time $t$ can be written in terms of the initial density matrix elements $\rho_{ij}$ ($i,j=1,2$) as follows
\eq
\rho_t =
 \begin{pmatrix}
 1-|G(t)|^2\rho_{22}&G(t)\rho_{12}\\
 \\
 G^*(t)\rho_{12}^*&|G(t)|^2\rho_{22}\\
\end{pmatrix}.
\eeq
The dynamics of a single amplitude damped qubit is captured by the time-local master equation \cite{bp}: 
\eq 
\frac{d\rho_t}{dt}=\gamma_1(t)\left[\sigma_{-}\rho_t\sigma_{+}-\frac{1}{2}\{\sigma_{+}\sigma_{-},\rho_t\}\right],
\eeq
where $\sigma_{\pm}$ are the spin lowering and rising operators and
\eq
\gamma_1(t)=-2\Re\frac{\dot{G}(t)}{G(t)}. \label{gammaADC}
\eeq
The function $G(t)$ depends on the form of the reservoir spectral density and is given in Appendix A for the two models here considered.

The state of the density matrix of the qubit at time $t$ can be written in terms of the initial density matrix elements $\rho_{ij}$ ($i,j=1,2$) as follows
\eq
\rho_t =
 \begin{pmatrix}
 1-|G(t)|^2\rho_{22}&G(t)\rho_{12}\\
 \\
 G^*(t)\rho_{12}^*&|G(t)|^2\rho_{22}\\
\end{pmatrix}.
\eeq

\subsection{Single qubit: the measures}

{\it RHP Measure} 

Since the master equation is in Lindblad form with time-dependent coefficients it is straightforward to evaluate the RHP measure for a generic spectral density \cite{EM}.
  \eq
\mathcal{N}_{\text{RHP}}=-\int_{\gamma_1(t)<0}dt\:\gamma_1(t).
  \eeq

We consider first the case of a Lorentzian spectrum 
\eq
J(\omega)=\frac{\gamma_M\lambda^2}{2\pi[(\omega-\omega_c)^2+\lambda^2]},
\eeq
with $\gamma_M$ an effective coupling constant, $\lambda$ the width of the Lorentzian and $\omega_c$ the peak frequency. When the qubit frequency, denoted $\omega_0$, coincides with $\omega_c$  (resonant Jaynes-Cummings model), the dynamical map is non-divisible for $r>r_{\text{crit}}=0.5$, with $r=\gamma_M/\lambda$ \cite{bp}. From this critical value $\mathcal{N}_{\text{RHP}}$ diverges as a direct consequence of the divergent behavior of $\gamma_1(t)$. Conversely, in the weak coupling regime, i.e. $r<0.5$, $\gamma_1(t)$ is positive for all times and hence the channel is always divisible ($\mathcal{N}_{\text{RHP}}=0$).

For the PBG model \cite{PBG}, using Eqs. (\ref{gammaADC}) and (\ref{gPBG}) we can study the Markovian to non-Markovian crossover as a function of the reservoir parameter $z=\Delta_P/\beta$, with $\Delta_P$ the detuning of the qubit frequency from the edge frequency $\omega_e$ of the band gap spectrum,  and $\beta$ a characteristic frequency.  Positive values of $z$ correspond to the case in which the qubit is outside the band gap region while negative values of $z$ correspond to the qubit in the band gap region. In the latter case the well-known phenomenon of population trapping occurs as the emission of energy in the reservoir is strongly inhibited.

For $z<z_{\text{crit}}=1.7$,  the rate $\gamma_1 (t)$ temporarily attains negative values for certain time intervals. In fact, due to population trapping, for $z \ll z_{\text{crit}}$ the asymptotic long time limit is characterized by small amplitude oscillations between positive and negative values which persist as $t \rightarrow \infty$. This eventually leads to a divergency not only of the $\mathcal{N}_{\text{RHP}}$ measure but of all non-Markovianity measures. In any practical experimental situation, however, the time of the experiment is finite. We will therefore calculate the measures for a fixed time interval, longer compared to the typical times of the system but, of course, shorter than the system-reservoir correlation time which in this case is $\infty$.

In Fig. \ref{AD} ii) we plot the RHP measure (red stars) for the PBG reservoir as a function of the parameter $z$. As we can see from the plot the measure has a sudden peak at values of $z$  close to the edge $z=0$, reaching its maximum value for $z=1.0$ before vanishing for $z=1.7$. For increasingly negative values of $z$ under the critical point, RHP non-Markovianity measure decreases to small but finite values. This is due to the decreasing amplitude of the oscillations in the decay rate.

{\it BLP Measure } 

We begin by deriving the analytical expression for $\mathcal{N}_{\text{BLP}}$ \cite{EM}: 
\eq
\mathcal{N}_{\text{BLP}}=- \max_{m,n} \int_{\gamma_1<0}dt\:\gamma_1(t)\frac{|G(t)|^3m^2+0.5|G(t)||n|^2}{\sqrt{|G(t)|^2m^2+|n|^2}}.
\eeq
where $m=\rho_{11}^1(0)-\rho_{11}^2(0)$ and $n=\rho_{12}^1(0)-\rho_{12}^2(0)$ are coefficients to be optimized. 
We have compelling numerical evidence that the maximizing states are the orthogonal states $\ket{+}\bra{+}$ and $\ket{-}\bra{-}$ for both the Lorentzian and the PBG spectral densities for any time $t$.  Hence the BLP measure takes the form
\eq
\mathcal{N}_{\text{BLP}}=-\frac{1}{2}\int_{\gamma_1<0}dt\:\gamma_1(t)|G(t)|. \label{NBLPADC}
\eeq
As we expect when only one decay rate is present in the master equation, $\mathcal{N}_{\text{BLP}}\neq 0$ if and only if the dynamics are non-divisible i.e., $\gamma_1(t)<0$. 

Figure \ref{AD} shows $\mathcal{N}_{\text{BLP}}$ (blue asterisk) for different values of i) $r$ and ii) $z$ for the Lorentzian and PBG spectra, respectively. The behavior is qualitatively similar to the one of the RHP measure. In the PBG case the peak of non-Markovianity is slightly shifted towards more negative values of $z$. In Ref. \cite{Rug}, the BLP measure and RHP witness are calculated for quantum harmonic oscillators in a band gap showing that both measures are sensitive to the edge of the gap, which is what we also observe.

%We mention that the authors in Ref. \cite{JCmodel} confirm the optimal states they find for this case are not correct and indeed, do not satisfy the necessary requirements proven to maximize the measure \cite{Optimal}.

%\cite{Cap_GioFaz, Cap_BAF} \cite{Cap_GioFaz, Cap_BAF}
{\it BCM and LFS Measures } 

For the amplitude damping channel the quantum and entanglement-assisted classical capacities, which we indicate here with $Q^A$ and $C_{ea}^A$, respectively, are calculated numerically \cite{num1, num2}. The states optimizing $I_c(\rho_t, \Phi_t)$ and $I(\rho_t, \Phi_t)$ are now time-dependent. One finds \cite{CQAD} the following formulas $C_{ea}^A  = \max_{p\in [0,1]} \Big\{ H_2(p) + H_2(|G(t)|^2p) -   H_2([1-|G(t)|^2]p)\Big\}$,
and $Q^A = \max_{p\in [0,1]} \Big\{ H_2(|G(t)|^2p) -   H_2([1-|G(t)|^2]p)\Big\}$, which still need a simple optimization over the probability $p\in [0,1]$. 
The latter formula holds only for $|G(t)|^2 >  \frac 12$, otherwise $Q(\Phi^A_t)\equiv 0$. This is due to the fact that the amplitude damping channel is  degradable for $|G(t)|^2>\frac{1}{2}$, while for $|G(t)|^2 \le \frac{1}{2}$ is anti-degradable with zero quantum capacity. 

The behaviour of the BCM and LFS measures in the two cases of amplitude damping channels is illustrated in Fig. \ref{AD}.
For both the Lorentzian reservoir spectrum and the PBG the measures $\N_{\text{C}}$ and $\N_\text{I}$ take non-zero values if and only if the amplitude damping channel is non-divisible. Notice that the measures $\N_\text{I}$ and $\N_\text{C}$ have very close values. This may seem not surprising given that both measures are based on quantum mutual information. However, the examples show that there is no relation between them even in the simple amplitude damping model here considered. Indeed a strong dependence on the form of the environmental spectrum can be noticed. More precisely, in the case of the Lorentzian spectrum we have $\N_\text{C}(\Phi^A)>\N_\text{I}(\Phi^A)$, while in the PBG model the opposite relation holds (see Fig. \ref{AD} i) and \ref{AD} ii), respectively). 

We would like to emphasize the difference in the behavior of $\N_\text{Q}$ for the Lorentzian reservoir spectrum. As shown in Fig. \ref{AD} i), indeed, unlike the other measures  $\N_\text{Q}$ is equal to zero even for a non-divisible channel and detects non-Markovianity only in a very strong coupling regime, i.e. when $r>43$. This is due to the fact that the amplitude damping channel is anti-degradable for $|G(t)|^2 <\frac{1}{2}$ so, from a quantum information processing point of view, only revivals that occur in the region $|G(t)|^2>\frac{1}{2}$ are important. 

The above example is consistent with the intuitive idea that the transmission of quantum information along a quantum channel is more sensitive to noise than the transmission of classical information (although assisted by entanglement shared between Alice and Bob). Once again this conclusion is, however, spectrum-dependent. In the case of the PBG model (Fig. \ref{AD} ii)) it is possible to set the parameters such that the noise in the channel has almost the same effect on both kinds of information. This is possible for  $z<0$, because $|G(t)|^2$ in this regime oscillates only above the value $\frac{1}{2}$, but it is no longer true for $0\leq z < 2$ as shown in Fig. \ref{AD} ii) -- the biggest difference occurring for $z=0$.

%{\it LFS Measure \textcolor{blue}{(Bogna)}} In case of the Lorentzian reservoir spectrum it is straightforward to show that the decay rates $\gamma(t)$ are always positive in the weak coupling regime, i.e. for $R\le \frac{1}{2}$, while it can take temporarily negative values for $R > \frac{1}{2}$, making the dynamical map in this regime non-divisible. Numerical results indicate that measure $\N_I$ takes non-zero values whenever the channel is non-divisible. 

\subsection{Two qubits in independent environment: the model}

For two qubits interacting with identical non-correlated environments the time evolution can still be calculated analytically \cite{PRL}. The solution is given in Appendix A.
It is straightforward to confirm that, as for the pure dephasing case, the corresponding master equation can be written as the sum of two Lindblad-like terms, describing the dynamics of each qubit respectively, with time-dependent coefficent $\gamma_1(t)$ given by Eq. (\ref{gammaADC}).

{\it RHP Measure} 

Directly from the form of the master equation, we immediately can show that
  \eq
\mathcal{N}_{\text{RHP}}=-2\int_{\gamma_1(t)<0}dt\:\gamma_1(t).
\label{RR}
  \eeq
As one would expect the measure is additive and, hence, for the PBG model it behaves identically to the single qubit case of Fig. \ref{AD} ii). On the other hand, for the Lorentzian spectrum, $\mathcal{N}_{\text{RHP}}=\infty$ when the dynamical map is non-divisible as it is in the one qubit case.

{\it BLP Measure} 

We numerically prove that for the case of the Lorentzian spectrum, the maximizing pair is $\ket{\pm\pm}\bra{\pm\pm}$. In this case we obtain the following expression for the BLP measure
\eq
\mathcal{N}_{\text{BLP}}=-\int_{\gamma_1<0} dt\: \gamma_1(t)\frac{|G(t)|-2|G(t)|^3+1.5|G(t)|^5}{\sqrt{2-2|G(t)|^2+|G(t)|^4}}.
\eeq

This expression is clearly different from Eq. (\ref{NBLPADC}) for the single qubit case, and we can show that the measure is sub-additive in this case. However, its qualitative behavior as $r$ changes is exactly the same as the single-qubit case and the renormalized value of $\mathcal{N}_{\text{BLP}}$ gives exactly the same curve as the one shown in Fig. \ref{AD} i) (blue asterisk).

The Photonic Band Gap model presents a number of difficulties. Indeed in this case the pair of states maximizing the increase in trace distance is very strongly dependent on the time interval chosen. We numerically calculate the measure as a function of $z$ in Fig. \ref{A1} a) iii). The non-Markovianity measure corresponds to the highest value of each column of states. In more detail, from Fig. \ref{A1} a) iii) we see the measure is maximized for initial pairs of mixed and pure states (blue dots) for $-15\leq z\leq -3$, maximally entangled (purple dots) for $-2\leq z\leq -1$, pure (pink dots) for $z=0$ and the tensor product state $\ket{\pm}\bra{\pm}$ (yellow dots) for $z=1$. We have not been able to exactly identify the states for which the increase in trace distance is maximal, even for a fixed time interval. By comparing the numerical value of $\mathcal{N}_{\text{BLP}}$ with the single qubit case of Fig. \ref{AD} ii), we see however that also in this case the renormalized quantity has the same qualitative behavior for one and two qubits. In this case we have verified that the BLP is superadditive for $-15\leq z \leq -2$ and subadditive for $-1\leq z \leq 1$. Moreover, the measure is zero if and only if the channel is divisible.

{\it LFS Measure}

Extensive numerical optimization shows that the state maximizing the quantum mutual information in Eq. (\ref{NI}) for two independent identical amplitude channels is of the form $\rho_{AB}^*=\rho^*\otimes \rho^*$, where state $\rho^*$ is the maximizing state for the one-qubit channel discussed in Sec. V B. Hence, as for the two-qubit independent dephasing channels, the measure $\mathcal{N}_\text{I}$ is additive. This holds for both  the Lorentzian and the PBG spectral densities. Therefore, the two-qubits behavior of the measure for different values of $r$ or $z$ is exactly the same as the one shown in Fig. \ref{AD}.

{\it BCM Measures}

Having in mind that whenever the amplitude damping channel is not degradable it is anti-degradable, and hence has zero quantum capacity, one can clearly see that the measures $\mathcal{N}_\text{C}$ and $\mathcal{N}_\text{Q}$ are additive and therefore  display identical behavior with respect to the system parameters as the one discussed in the one qubit scenario, shown in Fig. \ref{AD}.

\section{Discussion and Conclusions}
 
Let us now discuss the comparison between the measures for the single and composite open quantum systems here considered. The first observation that appears evident when looking at Figs. \ref{OQ}, \ref{TQ}, and \ref{AD}  ii).  is that, generally, the behavior of the RHP measure is different from that of all the other measures. Indeed, after the crossover from Markovian to non-Markovian, this measure tends to present a monotonically increasing behavior, while the other measures often have a maximum, i.e. there exist values of the reservoir parameters for which the memory effects are maximal. This fact can be traced back to the very definition of RHP measure, which counts and sums the areas of negativity of the time-dependent decay rates in the master equation. Often, the more structured is the environment, the greater is the number of negativity intervals and therefore the bigger is the RHP measure. Physically, the number of reverse jumps is increasing too, leading to greater re-coherence. However, the measure proves problematic for the Lorentzian spectrum where, as a direct consequence of the decay rate diverging when the dynamics is non-divisible, the measure diverges.  Also, we notice that this measure is additive for independent reservoirs, and the the qualitative behavior for two qubits in either independent and common (dephasing) environment is the same. For the other measures the situation is not so straightforward, as we explain below.

We start from the pure dephasing cases, Fig. \ref{OQ} (single qubit and two qubits in independent environment) and Fig. \ref{TQ} (2 qubits in common environment) show a clear similarity in the behavior of the BLP, LFS and BCM measures, in the sense that they all have a peak for values of $s$ between 2 and 4. This means that manifestations of memory in terms of increase of information on the system, increase of system-ancilla total correlations, or in terms of increase of channel capacities arise in a similar way when modifying the form of the spectrum. 

In the common environment case of Fig. \ref{TQ}, we note that, contrarily to the RHP measure, the other measures show a stronger sensitivity to the distance between the qubits, which in turn is connected to the cross-talk term, i.e. the environment mediated interaction that is known to contribute to the overall memory effects \cite{Laura}. The BLP, LFS and BCM measures seem to show a narrower peak as the distance is increased, consistently with the independent qubit case of Fig. \ref{OQ}.

The amplitude damping case presents clear differences and, contrarily to the pure dephasing case, the crossover between Markovian and non-Markovian is not the same for all the measures. We note first of all that the presence of energy exchange between the system and the environment introduces a new relevant time scale, or frequency, i.e. the Bohr frequency $\omega_0$ of the two-level system forming the qubit. In the dephasing case the structure of the spectral density and the presence of peaks in resonance with $\omega_0$ is not related to the occurrence of non-Markovian dynamics. Rather, it is the form of the spectrum at the origin $\omega=0$ that dictates the presence or not of re-coherence and revivals of information \cite{Timediscord}. 

The situation is clearly different in the dissipative case where the qubit is more likely to interact with environmental modes of the same frequency of $\omega_0$. A clear sign of this behavior is shown in Fig. \ref{AD} ii) for the Photonic Band Gap case. Here all non-Markovianity indicators display the same key feature, i.e., they have their maximum around $\omega_0 = \omega_e$, where the coupling between the qubit and the modes is the strongest \cite{Barry}. In this case, indeed, the qubit exchanges periodically energy with the environment and its population shows Rabi oscillations. Consistently, memory effects associated to the energy exchange between system and environment also lead to oscillations of the information content of the system (back flow of information), total correlations between system and ancilla, and channel capacities.

Finally, another important point to notice emerges from the comparison of the (finite) measures for the Lorentzian model on resonance. Here the behavior of the BLP and entanglement-assisted capacity is similar while for the quantum channel capacity measure, a much stronger coupling with the environment is required to have a partial increase in the maximum rate of information transfer for increasing times or lengths of the channel. As we have discussed in Sec. V B, this is not surprising as quantum information is more sensitive to environmental noise than classical information.

The overall picture that surfaces is one in which, despite the obvious differences between the measures, their corresponding physical mechanisms contributing to memory effects often appear correlated and show a similar connection with the reservoir spectral features. 
%We wish to conclude and summarize this long comparative study by recalling the old Indian tale of the  \lq\lq Blind Men and the Elephant\rq \rq. The story illustrates how perception is based on what a person is able to see or touch. In our case the elephant is non-Markovian dynamics and the blind men are the different measures.  Although each man touches the same animal, his determination of the elephant is based only on what he is able to perceive or, in our terms, on what physical phenomenon stemming from reservoir memory the measure is revealing. 
In conclusion the non-Markovianity measures give different perspectives on the same complex physical process, a full understanding of which requires them all.

\section{Acknowledgements}
B.B. thanks the Open Quantum Systems and Entanglement group for the hospitality, and acknowledges financial support form the Finnish Emil Aaltonen Foundation and the National Science Center project 2011/01/N/ST2/00393. D. C. was partially supported by the National Science Center project DEC-2011/03/B/ST2/00136. C.A. acknowledges financial support from the EPSRC (UK) via the Doctoral Training Centre in Condensed Matter Physics. The authors would also like to gratefully acknowledge Oliver Brown for his assistance in running numerical simulations using his computer. 

\section{Appendix}

\begin{figure*}[htp]
\centering
\includegraphics[width=0.8\textwidth]{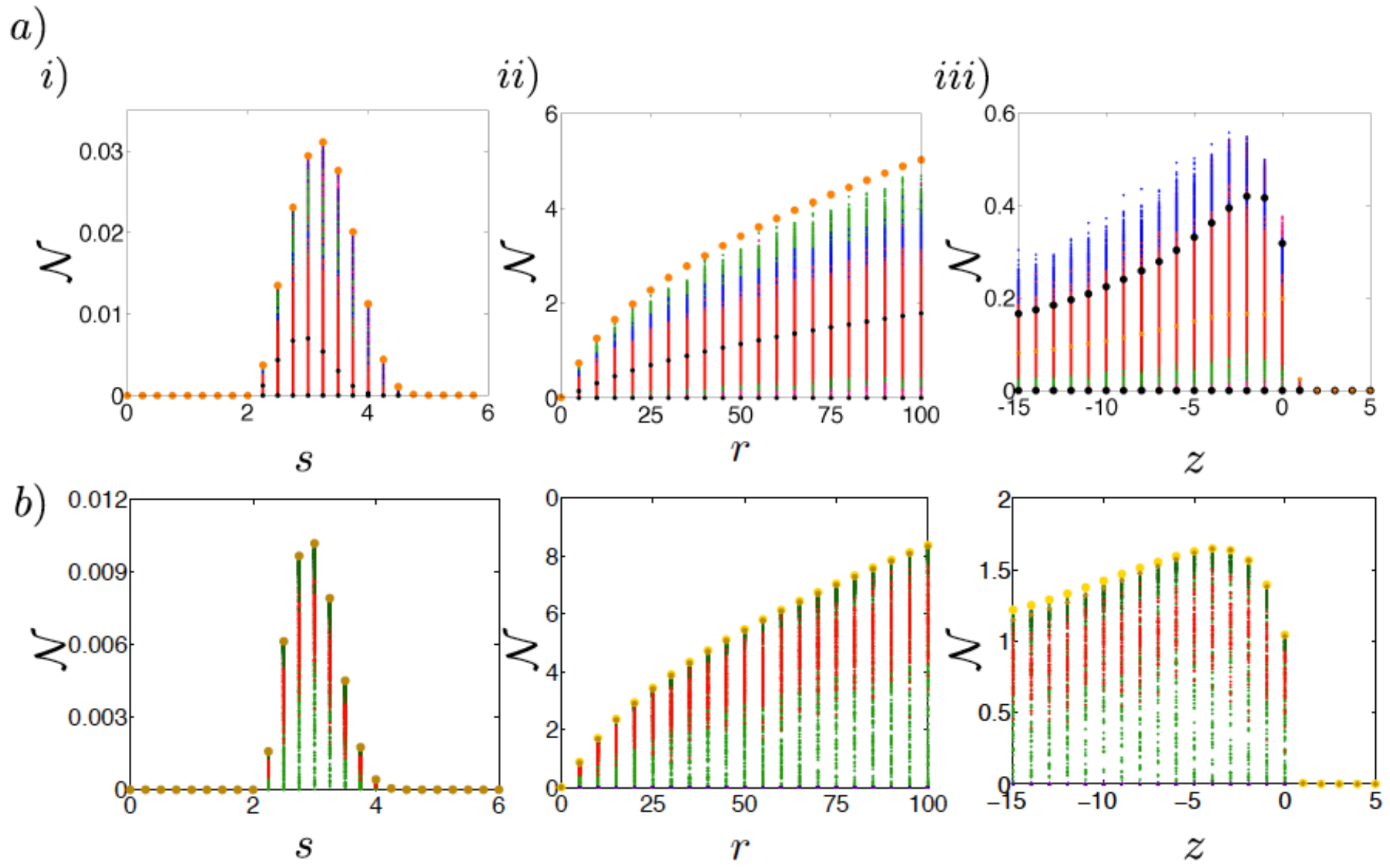}
\caption{(Color Online) Non-Markovianity Measure (the largest value in each column for each plot) for independent two qubit systems for the following reservoir spectra: i) Ohmic, ii) Lorentzian and iii) Photonic Band Gap model. We define the dynamics using a) the Breuer, Laine, Piilo Measure and b) the Luo, Fu, Song Measure.  All the measures are plotted against an environmental parameter which may be modifed. In general, to maximize each measure, random states are used, including maximally entangled (purple), pure (pink), mixed states (red) and product states (green). For the Breuer, Laine and Piilo measure we include combinations of mixed and pure states (blue), Bell states (black) and the tensor product state $\ket{\pm\pm}\bra{\pm\pm}$ (yellow). For all other measures we include separable states other than product states (dark green), the maximally mixed state (brown) and tensor product state of the optimizing states for the one qubit case (gold), which is parameter dependent. We consider for the Ohmic, Lorentzian and Photonic the following times periods: $t\in[0,20]$ in units of $\omega_c t$, $t\in[0,40]$ in units of $\lambda t$ and $t\in[0,20]$ in units of $\beta t$.}
\label{A1}
\end{figure*}

\subsection{Mathematical Description of Physical Models}
In this section, we present in detail the mathematical description of each system considered in this work. 
\subsubsection{Purely Dephasing Model} 
{\it One Qubit} The Hamiltonian of the system is given as \cite{puremodel2}: 
\eq
H=\omega_0\sigma_z+\sum_k\omega_k a^{\dagger}_ka_k+\sum_k\sigma_z(g_za_k+g^*_ka_k^{\dagger}),
\eeq
with $\omega_0$ the qubit frequency, $\omega_k$ the frequencies of the reservoir modes, $a_k(a_k^{\dagger})$ the annihilation (creation) operators of the bosonic environment and $g_k$ the coupling constant between each reservoir mode and the qubit. In the continuum limit $\sum_k|g_k|^2\rightarrow\int d\omega \: J(\omega) \delta (\omega_k-\omega)$, where $J(\omega)$ is the reservoir spectral density \cite{puremodel2, NMPure4}. 

It is simple to obtain the operator-sum representation $\phi^D_t(\rho)=\sum_{i=1}^2 K_i(t)\rho K_i^{\dagger}(t)$ with time-dependent Kraus operators; $K_1(t)=\sqrt{\frac{1+e^{-\Gamma}}{2}}\mathbb{I}$ and $K_2(t)=\sqrt{\frac{1-e^{-\Gamma}}{2}}\mathbb{I}$. 

Knowledge of the Kraus operators allows one to immediately also write the complementary map, needed to calculate both the coherent information and the entropy exchange which appears in the definition of the mutual information of the channel: 
\eqa
\tilde{\phi}_t^D[\rho]&=&\frac{1}{2}\left[(1+e^{-\Gamma(t)})\ket{1}_e\bra{1}+(1-e^{\Gamma(t)})\ket{2}_E\bra{2}\right]\nn\\&+&\frac{1}{2}\sqrt{1-e^{2\Gamma(t)}}\text{Tr}(\rho\sigma_z)\left(\ket{1}_E\bra{2}+\ket{2}_E\bra{1}\right).
\eeqa

We write in full Eq. \ref{gamma0} to give the explicit form of $\Gamma(t)$: 
\eqa
\Gamma(t)&=&\frac{2\tilde \Gamma[s]}{-1 + s}(1-(1+t^2)^{-s/2}(\cos(s\arctan(t))\nn\\&+&t\sin(s\arctan(t))).
\eeqa

{\it Two Qubit} The Hamiltonian which describes the two qubits $i,j$ for the purely dephasing case is as follows: \cite{puremodel2}: 
\eqa
H&=&\omega_0^i\sigma^i_z+\omega^i_0\sigma^j_z+\sum_k\omega_k a^{\dagger}_ka_k\nn\\&+&\sum_k\sigma^i_z(g^i_ka_k^{\dagger}+g^{i*}_ka_k)+\sum_k\sigma^j_z(g^j_ka_k^{\dagger}+g^{j*}_ka_k),
\eeqa

The expression for the ``cross-talk" term $\delta(t)$ is given below: 
\begin{widetext}
\begin{eqnarray}\label{gammapp}
\delta(t)&=&\frac{2\Gamma[s]}{-1 + s}(\{(1+ t_s^2)[1 + (t_s -t)^2]\}^{-\frac{s}{2} }\{[1 + (t_s- t)^2]^\frac{s}{2}\cos[s\text{arctan}(t_s)]+t_s [1+(t_s-t)^2]^\frac{s}{2}\sin[s\text{arctan}(q)]\nn\\&-&(1+t_s^2)^\frac{s}{2} (\cos[s\text{arctan}(t_s-t)]+(t_s-t)\sin[s\text{arctan}(t_s-t)])\}+\{(1+t_s^2)[1+(t_s+t)^2]\}^{-\frac{s}{2}}\{[1+(t_s+ t)^2]^\frac{s}{2}\nn\\& \times&\cos[s\text{arctan}(t_s)]\nn+t_s[1+(t_s+t)^2]^\frac{s}{2}\sin[s\text{arctan}(t_s)]-(1+t_s^2)^\frac{s}{2}[\cos[s\text{arctan}(t_s+t)]+(t_s+t)\sin\{s\text{arctan}(t_s+t)\}]\})
\end{eqnarray}
\end{widetext}

The form of the Kraus operators for this case is as follows 
$$K_1=\left(
       \begin{array}{cccc}
        e^{-\frac{1}{2}\Gamma_-}&0&0&0\\
  0&e^{-\frac{1}{2}\Gamma_+}&0&0\\
0&0&e^{-\frac{1}{2}\Gamma_+}&0\\
0&0&0&e^{-\frac{1}{2}\Gamma_-}
 \end{array}
     \right),
$$
$$K_2= (e^{-\Gamma_-}-1)\sqrt{e^{-\Gamma_-}+1}\left(
       \begin{array}{cccc}
      1 &0&0&0\\
  0&0&0&0\\
0&0&0&0\\
0&0&0&0
 \end{array}
     \right),
$$
$$K_3=\sqrt{1-e^{-\Gamma_-}}\left(
       \begin{array}{cccc}
        -e^{-\Gamma_-}&0&0&0\\
  0&0&0&0\\
0&0&0&0\\
0&0&0&1
 \end{array}
     \right),
$$
$$K_4=(e^{-\Gamma_+}-1)\sqrt{e^{-\Gamma_+}+1}\left(
       \begin{array}{cccc}
      0 &0&0&0\\
  0&1&0&0\\
0&0&0&0\\
0&0&0&0
 \end{array}
  \right),
$$
$$K_5=\sqrt{1-e^{-\Gamma_+}}\left(
       \begin{array}{cccc}
       0&0&0&0\\
  0& -e^{-\Gamma_+}&0&0\\
0&0&1&0\\
0&0&0&0
 \end{array}
     \right).$$

\subsubsection{Amplitude Damped Models} 
{\it One Qubit} The following microscopic Hamiltonian model describing a two-state system interacting with a bosonic quantum reservoir at zero temperature is given by \cite{bp}
\eq
H=\omega_o\sigma_z+\sum_k\omega_k a^{\dagger}_k a_k +\sum_k (g_k a_k\sigma_+ +g^*_k a_k^{\dagger}\sigma_-)
\eeq
As usual, $\sigma_{\pm}$ are standard raising and lowering operators respectively. 

The Kraus representation $\phi_t^A(\rho)=\sum_{i=1}^2K_i(t)\rho K_i^{\dagger}(t)$ for the amplitude damping channel is given by $K_1=\bigl(\begin{smallmatrix}1&0\\ 0&G\end{smallmatrix} \bigr)$ and $K_2=\bigl(\begin{smallmatrix}0&\sqrt{1-|G|^2}\\ 0&0\end{smallmatrix} \bigr)$ which gives us a complementary map defined by: 
\eqa
\tilde{\phi}_t^A(\rho)&=&[1-(1-|G(t)|^2)\rho_{22}]\ket{1}_E\bra{1}\nn\\&+&(1-|G(t)|^2)\rho_{22}\ket{2}_E\bra{2}\nn\\&+&\sqrt{1-|G(t)|^2}(\rho_{12}\ket{1}_E\bra{2}+\rho_{21}\ket{2}_E\bra{1}).
\eeqa
%IMPORTANT: PLEASE CHANGE HERE R WITH r AS WE HAVE USED R FOR THE DISTANCE BETWEEN QUBITS IN SEC IV

\begin{figure*}[htp]
\centering
\includegraphics[width=0.7 \textwidth]{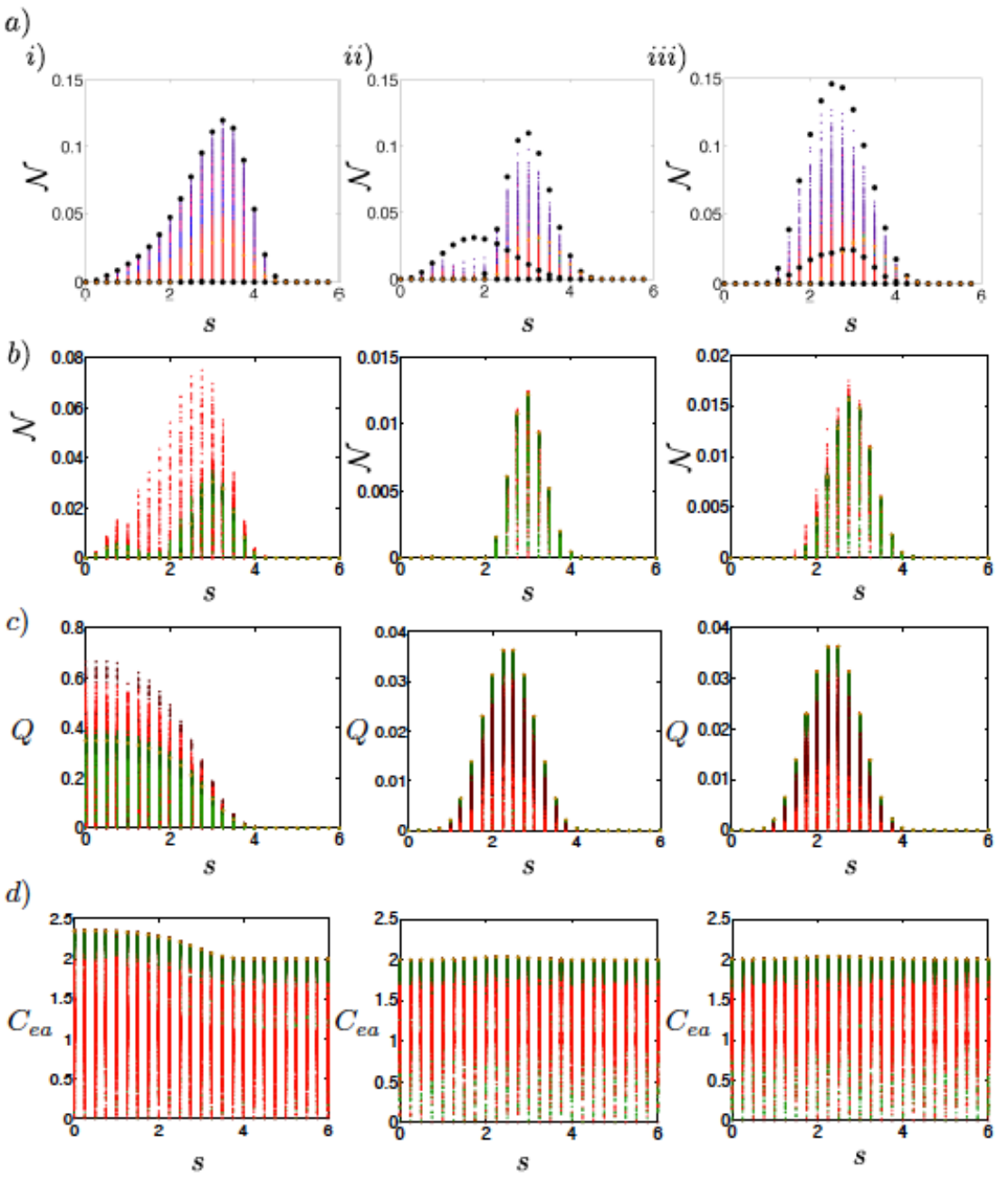}
\caption{(Color Online) Non-Markovianity Measure (the largest value in each column for each plot) for common two purely dephasing qubit systems for transit times $t_s$ i) 0.25, ii) 2 and iii) 6. We define the dynamics using a) the Breuer, Laine, Piilo Measure, b) the Luo, Fu, Song Measure, c) the Quantum Capacity Measure and d) the the Entanglement-Assisted Classical Capacity Measure.  All the measures are plotted against the Ohmic parameter $s$ which may be modified. In general, to maximize each measure, random states are used, including maximally entangled (purple), pure (pink), mixed states (red) and product states (green). For the Breuer, Laine and Piilo measure we include combinations of mixed and pure states (blue), Bell states (black) and the tensor product state $\ket{\pm\pm}\bra{\pm\pm}$ (yellow). For all other measures we include separable states other than product states (dark green), mixed states of rank 2 with eigenvalues close to $\frac{1}{2}$ (dark red), the maximally mixed state (brown). We consider a time interval $t\in[0,20]$ in units of $\omega_c t$. Note for rows c) and d) we optimize for $t=5$ for the Quantum Capacity $Q$ and the Entanglement-Assisted Classical Capacity  $C_{ea}$ respectively.  }
\label{A2}
\end{figure*}

We now discuss the specific forms of $G(t)$ for both amplitude damped systems, starting with the Lorentizan spectrum \cite{bp} using notation $G^L(t)$. 

If the spectral density has a Lorentzian shape, i.e, $J(\omega)=\gamma_M\lambda^2/2\pi[(\omega-\omega_c)^2+\lambda^2]$, then the function $G^L(t)$ takes the form: 
\eq
G^L(t)=e^{-\frac{(\lambda-i\Delta_L)t}{2}}\left[\text{cosh}\left(\frac{\Omega t}{2}\right)+\frac{\lambda-i\Delta_L}{\Omega}\text {sinh}\left(\frac{\Omega t}{2}\right)\right],
\eeq
with 
\eq
\Omega=\sqrt{\lambda^2-2i\Delta_L \lambda -4\varpi^2}
\eeq
where $\varpi=\gamma_M\lambda/2+\Delta_L^2/4$ and $\Delta_L=\omega_0-\omega_c$. 

For $\Delta_L=0$, one obtains the following solution
\eqa
G^L(t)&=&e^{-\lambda t/2}\left[\text{cosh}\left(\sqrt{1-2r}\frac{\lambda t}{2}\right)\right.\nn\\&+&\left.\frac{1}{\sqrt{1-2r}}\text{sinh}\left(\sqrt{1-2r}\frac{\lambda t}{2}\right)\right]
\eeqa
with $r=\gamma_M/\lambda$. 

For the Photonic Band Gap model, the specific form of $G^P(t)$ is as follows \cite{PBG}:

%IMPORTANT: PLEASE DEFINE HERE ALL SYMBOLS AND AVOID TO USE SYMBOLS THAT WE HAVE USED BEFORE WITH OTHER MEANING: $\delta$, $a_i$, ETC.

\eqa
G^P(t)&=&2v_1x_1e^{\beta x_1^2+i\Delta_P t}+v_2(x_2+|x_2|)e^{\beta x_2^2 t+i \Delta_P t}\nn\\&-&\sum_{j=1}^{3}v_j|x_j|[1-\Phi(\sqrt{\beta x^2_j t})]e^{\beta x^2_j t+i\Delta_P t}. \label{gPBG}
\eeqa
where $\Delta_P=\tilde\omega_0-\omega_e$ is the detuning from the band gap edge frequency $\omega_e$, set to equal zero as we consider only the resonant case and $\Phi(x)$ is the error function, whose series and asymptotic representations are given in Ref. \cite{PG}. In addition: 
\eqa
& & x_1=(A_++A_-)e^{i(\pi/4)},\nn\\& & 
x_2=(A_+e^{-i(\pi/6)}-A_-e^{i(\pi/6)})e^{-i(\pi/4),} \nn\\& & 
x_3=(A_+e^{i(\pi/6)}-A_-e^{-i(\pi/6)})e^{i(3\pi/4)},
\eeqa
\eq
A_{\pm}=\left[\frac{1}{2}\pm\frac{1}{2}\left[1+\frac{4}{27}\frac{\Delta_P^3}{\beta^3}\right]^{1/2}\right]^{1/3},
\eeq
\eq
v_1 = \frac{x_1}{(x_1-x_2)(x_1-x_3)} 
\eeq
\eq
v_2 = \frac{x_2}{(x_2-x_1)(x_2-x_3)},
\eeq
\eq
\beta^{3/2}=\tilde\omega_0^{7/2}d^2/6\pi\epsilon_0\hbar c^3.
\eeq
The coefficient $\beta$ is defined as the characteriztic frequency, $\epsilon_0$ the Coulomb constant and $d$ the atomic dipole moment. We have defined in our results $z=\Delta_P/\beta$. 

We note, that $G(t)$ satisfies the non-local equation $\dot{G}(t)=-\int_0^tf(t-t')G(t')dt'$ with initial condition $G(0)=1$, and f(t) is the reservoir correlation function wihich is related via the Fourer transform with a spectral density $J(\omega)$.

{\it Two Qubit}
For the amplitude damped channel, we consider the Hamiltonian: 
\eqa
H&=&\omega_o\sigma_z^A+\omega_o\sigma_z^B+\sum_k\omega_k a^{\dagger}_k a_k \nn\\&+&\sum_k (g_k a_k\sigma^A_+ +g^*_k a_k^{\dagger}\sigma^A_-)\sum_k (g_k a_k\sigma^B_+ +g^*_k a_k^{\dagger}\sigma^B_-)\nn\\& & 
\eeqa
The diagonal elements of the amplitude damping channel $\phi_t^{A,B}$ are written as follows \cite{TwoqG}:
\eqa
& &\rho_{11}(t)=|G(t)|^4\rho_{11}(0) \nn\\& &
\rho_{22}(t)=|G(t)|^2\rho_{11}(0)(1-|G(t)|^2)+\rho_{22}(0)|G(t)|^2 \nn\\& & 
\rho_{33}(t)=|G(t)|^2\rho_{11}(0)(1-|G(t)|^2)+\rho_{33}(0)|G(t)|^2 \nn\\ & & 
\rho_{44}(t)=1-(\rho_{11}(t)-\rho_{22}(t)-\rho_{33}(t))
\eeqa
For the off-diagonal elements, we have: 
\eqa 
& &\rho_{12}(t)=|G(t)|^2G(t)\rho_{12}(0) \nn\\ & & 
\rho_{13}(t)=|G(t)|^2G(t)\rho_{13}(0) \nn\\ & & 
\rho_{14}(t)=G(t)^2\rho_{14}(0) \nn\\ & & 
\rho_{23}(t)=|G(t)|^2\rho_{23}(0) \nn\\ & &
\rho_{24}(t)=\rho_{13}(0)G(t)(1-|G(t)|^2)+\rho_{24}(0)G(t)\nn\\& & 
\rho_{34}(t)=\rho_{12}(0)G(t)(1-|G(t)|^2)+\rho_{34}(0)G(t) \nn\\
\eeqa
Since this is model for two independent identical environments, the Kraus operators are just tensor products of Kraus operators of one qubit case, $K_{ij}=K_i \otimes K_j$, for $i,j=1,2$.
\subsection{Optimal State Pairs}
In this section, we provide numerical evidence to support the claims made in this work concerning the optimal states. We do not include the RHP measure in any case as it requires no optimization. For any previously unseen one qubit calculations, the states optimizing the measure are either straightforward to realize analytically or numerically through restricting the state space using necessary conditions. Indeed, it is known for the BLP measure, that it is sufficient to sample only the antipodal states on the surface of the Bloch sphere \cite{Optimal} and an analytical proof detemining the optimal pair also exists in Ref. \cite{ana}. 

In Fig. \ref{A1}, we consider the independent case for all models using the BLP and LFS measure only as it is simple to analytically conclude that the BCM measures are additive. For the BLP measure (see Fig. \ref{A1} a) i) and a) ii), we see that for both the Ohmic and Lorentzian spectra, the initial pairs of optimal states are the separable states $\ket{\pm}\bra{\pm}$. For the Photonic Band Gap model, the maximizing state is dependent on $z$. From Fig. \ref{A1} a) iii, we see the measure is maximized for initial pairs of mixed and pure states (blue dots) for $-15\leq z\leq -3$, maximally entangled (purple dots) for $-2\leq z\leq -1$, pure (pink dots) for $z=0$ and the tensor product state $\ket{\pm}\bra{\pm}$ (yellow dots) for $z=1$.

For the LFS measure, for all models of the independent identical environments (see Fig. \ref{A1} b)) the optimizing states are in form of the product states of the state that optimizes the measure for one qubit dynamics. For the case of pure dephasing these are maximally mixed states (brown) and for both amplitude damping cases the states (gold) depend on the parameters characterizing the reservoir.  

In Fig. \ref{A2}, we consider the Ohmic case for all measures which require optimization for three different transit times $t_s$. For the BLP measure (see Fig. \ref{A2} a)), we see the optimal state is always an orthogonal pair of Bell states. In more detail, depending on the value of the Ohmic parameter $s$, the maximizing pair is either the sub- or super-decoherent Bell State pair $\ket{\Phi_{\pm}}\bra{\Phi_{\pm}}$ or $\ket{\Psi_{\pm}}\bra{\Psi_{\pm}}$. 

Optimization of the LFS measure in the case of common environments is much more challenging. In case where both qubits are closeby (see Fig. \ref{A2} b) i)) the optimizing states are of rank two with the eigenvalues given by $\lambda_{1,2}=\frac{1}{2}\pm \epsilon$, where $\epsilon \in [0, 0.1]$ (dark red), while for both cases of $t_s=2$ and $6$ (see Fig. \ref{A2} b) ii) and b) iii) ) optimizing states become close to maximally mixed states (with eigenvalues $\lambda_{1,2,3,4}=\frac{1}{4}+ \epsilon_{1,2,3,4}$, where $\epsilon_{1,2,3,4}$, $\in[-0.1,0.1]$ are such that the normalization condition is satisfied), that are the optimizing states in the independent environments model.

Unlike for for the independent case, here the additivity property for BCM measures is no longer valid, hence the optimization is needed. Notice that in this cases we are not optimizing measures $\N_\text{C}$ and $\N_\text{Q}$ directly, but rather looking for the states that give the channel capacities, $C_{ea}$ and $Q$ respectively. For $t_s=0.25$ for both capacities the optimizing states are from the same class as in the case of the LFS measure (dark red), for $t_s=2$ and $6$ the optimizing states are the maximally mixed states as in the independent environment scenario.

We would like to emphasise the clear effect of the ``cross-talk" term in the model of common reservoir dephasing on the measures LFS and BCM, that is reflected in the states that are optimizing the relevant quantities. The bigger the transit time gets, or equivalently the further the two qubits are apart form each other, the closer the model is to the independent environment case and hence the closer the optimizing states become to the maximally mixed state.

%\blue{[[BOGNA: Please add a few lines about your measures and add anything else I have missed. ]]}

\end{document}